\documentclass[twocol]{ametsoc}
\usepackage{mathtools, cuted}

\journal{jpo}

\bibpunct{(}{)}{;}{a}{}{,}

\title{The impact of locally-enhanced vertical diffusivity on  the cross-shelf transport of tracers induced by a submarine canyon}

\authors{Karina Ramos-Musalem\correspondingauthor{Department of Earth, Ocean and Atmospheric Sciences, University of British Columbia, 2020 – 2207 Main Mall, Vancouver, BC, Canada, V6T 1Z4} and Susan E. Allen}

\affiliation{Department of Earth, Ocean and Atmospheric Sciences, University of British Columbia, Vancouver, BC, Canada.}

\email{kramosmu@eoas.ubc.ca}

\abstract{The exchanges of water, nutrients and oxygen between the coastal and open ocean are key components of on-shelf nutrient budgets and biogeochemical cycles. On a regional scale, submarine canyons enhance physical processes such as shelf-slope mass exchange and mixing. There is good understanding of the flow around upwelling submarine canyons; however, the flux of biologically relevant tracers is less understood. This work investigates the impact of submarine canyons on the cross-shelf exchange of tracers and water, taking into account the impact of locally-enhanced mixing within the canyon, and develops a scaling estimate for canyon-induced upwelling of tracers, proportional to local concentration, vertical diffusivity, and previously scaled upwelling flux. For that purpose, we performed numerical experiments simulating an upwelling event near an idealized canyon, adding a passive tracer with an initially linear profile. We varied the geographic distribution of vertical eddy diffusivity and its magnitude, the initial stratification, Coriolis parameter, and the strength of the incoming flow. We find that a canyon, of width 5\% of the alongshelf length of the shelf, upwells between 25 to 89\% more tracer mass onto the shelf than shelf-break  upwelling. Locally-enhanced vertical diffusivity has a positive effect on the tracer that is advected by the upwelling flow and can increase canyon-upwelled tracer flux by up to 27\%.
}

\begin{document}

\maketitle

%








\section{ Introduction }
Exchange of water and solutes between the coastal and open ocean is key to understanding global biogeochemical budgets and their response to climate change and human activities \citep{Jordi2008}. Moreover, the specific spatial distribution of tracers like dissolved oxygen can impact benthic and demersal communities \citep{Keller2010}. In general, exchange between the deep ocean and the continental shelf is limited as homogeneous, geostrophic flow is restricted to follow isobaths along the continental shelf (Taylor-Proudman Theorem), such that deep ocean exchange occurs only when ageostrophic dynamics occur \citep{Allen2009}. Submarine canyons can induce ageostrophic motions because the canyon is a region of higher Rossby number relative to the slope, meaning that near the canyon advection of momentum is an important driver of the flow. 
On a regional scale, submarine canyons are known to modify or enhance shelf-slope mass exchange and regional currents \citep{Hickey1995}. 
\begin{figure*}[th!]
\begin{center}
   	\noindent \includegraphics[width=0.8\textwidth]{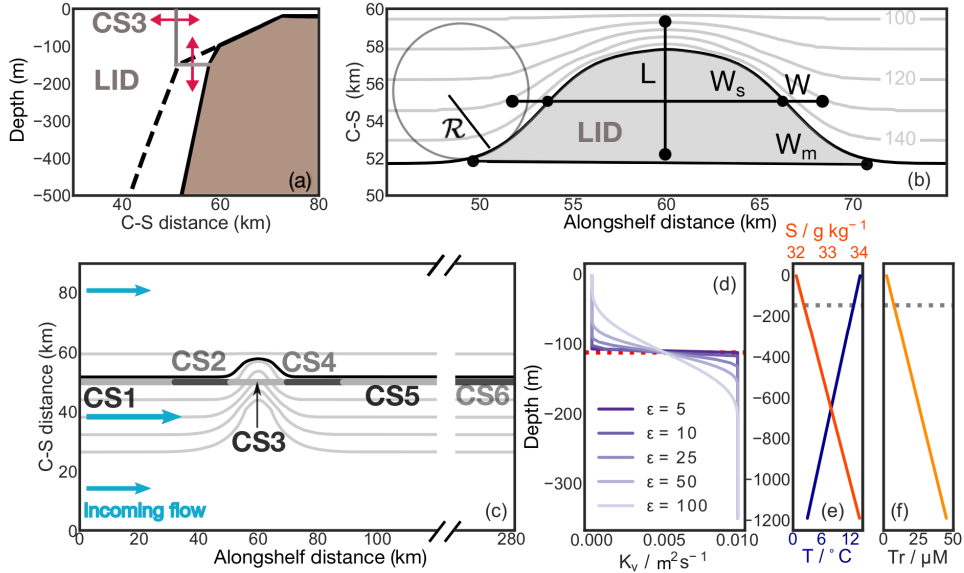} 
	\caption{(a) Cross-shore section through the canyon axis, the dashed line marks the shelf bottom, which can be identified with the canyon rim. (b) Top view of the canyon. The shaded area corresponds to the LID section across which vertical transport was calculated. The solid black line is the shelf-break at 149.5~m. Canyon dimensions: $L=8.3$~km is the length along the axis, $W_s=12.3$~km is the width at mid length at the shelf-break isobath, $W=21.1$~km is the width at mid-length at rim depth, $W_m=24.4$~km is the width at the mouth at shelf-break depth, and $\mathcal{R}=5.5$~km is the radius of curvature of the shelf-break isobath, upstream of the canyon. (c) Top view of the domain. The shelf volume is bounded by the wall that goes from shelf-break (black contour) to surface in the no-canyon case (sections CS1-CS6), alongshelf wall at northern boundary and cross-shelf walls at east and west boundaries. (d) Example of initial vertical diffusivity profiles at a station in the canyon with $K_{can}=10^{-2}$ ~m$^2$s$^{-1}$ and different values of $\epsilon$. Note that $\epsilon=5$~m corresponds to a step profile. The dashed red line indicates the depth of the canyon rim. Initial profiles of (e) temperature and salinity for the base case, and (f) tracer concentration with maximum and minimum values of 45 and 2 $\mu$M, respectively, and shelf-break depth concentration $C_s=7.2$ $\mu$M. The dashed gray line corresponds to the shelf-break depth.}
    \label{fig:sections}
    \end{center}
\end{figure*}
Both the distribution and on-shelf inventory of nutrients and oxygen can have relevant biological consequences for the shelf system. A recent numerical study of the coast of Washington State estimated that seasonal nitrate input from the slope to the shelf associated with three nearby canyons was between 30 to 60\% of that contributed by local wind-driven upwelling \citep{Connolly2014}. They also found that changes in near-shelf bottom oxygen concentrations in the presence of the canyons matched levels of hypoxia in the region. These changes were large enough to have an ecological impact if compared to levels of severe hypoxia associated with mortality in marine organisms. Moreover, it has been reported that on the west coast of the United States small changes in dissolved oxygen concentrations in already hypoxic waters can cause large changes in the total and species-specific catch of demersal fish \citep{Keller2017}.

In addition to enhancing upwelling, submarine canyons can enhance mixing within their walls by focusing internal waves and tides \citep{Gordon1976}. Although turbulence has been measured in only a few canyons, average diapycnal diffusivity values in the surveyed ones are very high compared to levels outside (e.g. Monterey Canyon $K_{D}\approx 2.5\times 10^{-2}$ m$^2$s$^{-1}$ \citep{Carter2002}, Ascension Canyon $K_{D} \approx 3.9\times10^{-3} $m$^2$s$^{-1}$ \citep{Gregg2011}, Gaoping Canyon $K_{D} \approx 10^{-2}$ m$^2$s$^{-1}$ \citep{Lee2009}, dissipation rates equivalent to $K_{D}\approx10^{-3}$m$^2$s$^{-1}$ at Eel Canyon \citep{Waterhouse2017}), so it is reasonable to assume that locally-enhanced mixing is a characteristic of many submarine canyons. 

There has been extensive research on the upwelling circulation within submarine canyons (\emph{eg}. \cite{Allen2010} hereafter AH2010, \cite{Howatt2013} hereafter HA2013, \cite{Freeland1982}, \cite{Klinck1996}, \cite{Kampf2007}). However, the slope-shelf flux of biologically relevant tracers, such as oxygen or nitrate is less understood. The objective of this work is to study the impact of an idealized submarine canyon on the cross-shelf exchange and on-shelf distribution of a passive tracer, taking into account the effect of locally-enhanced mixing. We quantify cross-shelf fluxes of a passive tracer and develop scaling estimates for the tracer flux upwelled onto the shelf. To quantify the impact of locally enhanced mixing we have designed numerical experiments that represent mixing in the form of enhanced vertical diffusivity and viscosity using different geographical distributions. 

In the following sections we explain the numerical configuration and experiments (Sec. \ref{sec:Methodology}); we describe the flow dynamics of the base case (Sec. \ref{sec:results}\ref{sec:results_flow}) and the effect of locally enhanced diffusivity on the dynamics of the flow and tracer transport from the canyon to the shelf; we look at the tracer evolution within the canyon (Sec. \ref{sec:results}\ref{sec:isopyc_isotracer}), cross-shelf transports (Sec.\ref{sec:results}\ref{sec:CStrans}) and upwelling through the canyon (Sec.\ref{sec:results}\ref{sec:HCW}). In section \ref{sec:scaling} we scale the advection-diffusion equation and provide justification for choosing the parameter space we explored in the numerical experiments. Furthermore, we develop a scaling estimate for the tracer flux onto the shelf as the product of a characteristic concentration and the canyon-upwelled water flux derived in previous scaling estimates (AH2010, HA2013) with a modification to account for the effect of enhanced mixing. Finally, in section \ref{sec:Discussion}, we provide a summary and discussion of our results.

\section{Methodology}
\label{sec:Methodology}
We use the \emph{Massachusetts Institute of Technology general circulation model} (MITgcm) \citep{Marshall1997} to simulate a system consisting of a sloping continental shelf cut by a submarine canyon (Fig.\ref{fig:sections}), with incoming flow from the west (upwelling favourable), parallel to the shelf. The range of stratifications, incoming shelf currents and Coriolis parameters selected for all runs represent realistic oceanic conditions over continental shelves around the world and, in this sense, they constitute typical dynamical settings for a submarine canyon. We explore a wider range of parameter space for vertical mixing, as we go from low values of vertical diffusivity to the extreme values observed both in magnitude and vertical distribution.
 
The simulation starts from rest. A shelf current is spun-up by applying a body force on every cell of the domain directed westward, alongshelf with a similar effect as changing the rotation rate of a rotating table \citep{Spurgin2014}. The body forcing ramps up linearly during the first day, stays constant for another day, and ramps down to a minimum, after which it remains constant and just enough to avoid the spin-down of the shelf current. This forcing generates a deeper shelf current, less focused on the surface, than the coastal jet generated by wind-forced models (SI, Fig. S1). The model was run for 9 days.
 
The domain is 280 km alongshelf and 90 km across-shelf divided in 616x360 cells horizontally. The cell width increases smoothly alongshelf and cross-shelf, from 115 m over the canyon to 437 m at the north boundary, and to 630 m at a distance of 60 km upstream and downstream of the canyon and then is uniform to the downstream boundary. Vertically, the domain is divided in 90 z-levels spanning 1200 m, with grid sizes varying smoothly from 5 m (surface to below shelf) to 20 m at depth. The time step used was 40 s, with no distinction between baroclinic and barotropic time steps. The experiments ran in hydrostatic mode. Some runs were also repeated in non-hydrostatic mode with no significant differences in the results.

The canyon was constructed from a hyperbolic tangent function. Geometric parameters of the canyon (Fig. 1b) are similar to those of Barkley or Quinault Canyons, with geometric and dynamical non-dimensional groups representative of numerous short canyons, as will be discussed in Section 4e (AH2010, \cite{Allen2000} for short canyon discussion). The domain has open boundaries at the coast (north) and deep ocean (south). Open boundaries use Orlanski radiation conditions and no sponge. Bottom boundary conditions are free-slip with a quadratic drag with coefficient 0.002, while vertical walls on the bathymetry steps have a free-slip condition. East and west boundaries are periodic. The domain is long enough that water does not recirculate through the canyon during the simulation. However, barotropic Kelvin waves, first and second mode baroclinic Kelvin waves, and long wavelength shelf waves do recirculate through the domain as in previous studies with similar configurations (e.g. \cite{She2000}, \cite{Dinniman2002}). Subinertial shelf-waves of wavelength likely to be excited by the canyon (40 km) \citep{Zhang2017} are too slow to recirculate. The gravest mode has a wave speed of approximately 0.5~ms$^{-1}$ against the flow (Calculated using \cite{Brink2006}). 

The initial fields of temperature and salinity vary linearly in the vertical and are horizontally homogeneous (Fig. \ref{fig:sections}e). For all runs, temperature decreases and salinity increases with depth but their maximum and minimum values are changed to generate the different stratifications used in the simulations. A passive tracer was introduced from the beginning of the simulations with a linear vertical profile that increases with depth, intended to mimic a nutrient such as nitrate (Fig. \ref{fig:sections}f). The maximum and minimum values of the profile come from data collected during the Pathways Cruise in summer, 2013 in Barkley Canyon \citep{Klymak2013}. 

We use the GMREDI package included in MITgcm for diffusing tracers. Since the mesoscale eddy field is resolved, we have no need to characterize the transport due to these structures. However, it is desirable to numerically handle the effects of tilting isopycnals that are intrinsic to canyon upwelling dynamics \citep{Allen2001}. Mixing and stirring processes are better described within the canyon as being along-isopycnal and cross-isopycnal, rather than horizontal and vertical. Inside the canyon vertical mixing is set to be larger than outside (see below), so diapycnal tracer transport will be enhanced. Considering this, we use the scheme for isopycnal diffusion \citep{Redi1982} but did not use the skew-flux parametrization \citep{Gent1990}. In sum, the vertical effective diffusivity on the tracer is determined by the prescribed vertical eddy diffusivity $K_v$, the tilting of isopycnals via the Redi scheme (vertical contribution) and the diffusivity due to the advection scheme, which is a 3rd order, flux-limited scheme that treats space and time discretizations together (direct space time) and uses non-linear interpolation (non-linear scheme) \citep{MITgcm}.

Patterns of enhanced diapycnal mixing within submarine canyons vary spatially and temporally. For example, Ascension Canyon, has sides and axis slopes supercritical to M2 internal tides, with maximum dissipation zones near the bottom, just below the rim, and larger average dissipation rates during spring tides \citep{Gregg2011}. On the other hand, Gaoping Canyon is subject to strong barotropic and baroclinic (1st mode) tides and, at critical frequencies, there is a turbulent overturning due to shear instability and breaking of internal tides and waves; diapycnal diffusivity varies seasonally due to changes in stratification \citep{Lee2009}. 

Diapycnal diffusivity profiles along Ascension Canyon's axis show a sharp gradient near rim depth, close to the head, and the mean profile shows a clear step in diffusivity at rim depth (SI, Figure S2, bottom row). We also see sharp but less intense variations of diapycnal diffusivity $K_{D}$ near the rim in the mean profile for Eel Canyon (SI, Figure S2, bottom row). Monterey Canyon also shows larger levels of diffusivity within the canyon, although the increase at the rim is less sharp than in Eel Canyon (SI, Figure S2, bottom row). 

Given that there is a wide range of mixing patterns in canyons, we developed a simple representation of enhanced vertical mixing within the canyon by prescribing higher values of $K_{v}$ and $A_v$ to grid points within the canyon ($K_{can}$) than background ($K_{bg}$) so that the profile of $K_{v}$ decreases sharply above the rim at every point along the canyon (Fig. \ref{fig:sections}d). We defined cells within the canyon as the residual grid points from subtracting the bathymetry of a straight shelf from one incised by a canyon.  Additionally, we ran experiments with  smoother $K_v$ profiles defined by the Heaviside function with a smoothing parameter $\epsilon$ 
\begin{strip}
\begin{equation}
K_v(z) = \begin{cases}
K_{bg} &\mbox{if  } z > H_r+\epsilon \\
K_{bg}+ K_{can}\left[0.5 + \frac{H_r-z}{2\epsilon} + \frac{1}{2\pi}\sin{\frac{\pi(H_r-z)}{\epsilon}}\right] & \mbox{if  } {H_r}-\epsilon < z < {H_r}+\epsilon \\
K_{can}+K_{bg} & \mbox{if  }  z < {H_r}-\epsilon
\end{cases} 
\label{eq:heaviside}
\end{equation}
\end{strip}
where $z$ is depth and $H_r$ is the rim depth. We define the rim depth at a point ($x_c$, $y_c$) within the canyon as the depth of the shelf away from the canyon, at cross-shelf distance $y_c$. Given the vertical resolution of our model, the smallest effective $\epsilon$ is 5 m. The length $\epsilon$ defines the smoothing length of the step, so the larger $\epsilon$ is, the smoother the profile. 

Upwelled water on the shelf has been estimated previously by finding water originally below shelf-break depth based on its salinity (HA2013). We take the same approach but use the tracer concentration at shelf-break depth as the criterion to find water on shelf that was originally below shelf-break depth. Enhanced diffusion may cause our algorithm to underestimate the amount of upwelled water on shelf. To minimize this error we added a second tracer with the same linear profile as the original but with smaller explicit diffusivity. This allows us to find upwelled water on the shelf without the effects of enhanced diffusivity on concentration, only keeping the dynamical effects of enhanced $K_{v}$ through modifications of density. The constant gradient of the linear profile also contributes to lower the numerical diffusivity compared to other profiles. The linear advective tracer is only used to find the upwelled water; all tracer mass integrations are over the original tracer with the mixing characteristics reported in Table \ref{tab:experiments_mix}. 

We explore the effects of vertical eddy diffusivity $K_{v}$ and vertical eddy viscosity $A_v$, locally-enhanced vertical diffusivity $K_{can}$ and viscosity  $A_{can}$, stratification $N_0$,  Coriolis parameter $f$, and incoming velocity $U$. All reported experiments (Tables \ref{tab:experiments} and \ref{tab:experiments_mix}) have $K_{v} = A_v$ since we found that the effect of $A_v$ is not significant and will not be discussed further. We report the effects of modifying $N_0$ and $f$ combined as the Burger number $B_u=N_0H_s(fW)^{-1}$, where $W$ is the width at mid-length at rim depth (Fig. \ref{fig:sections}b), and of $f$ and $U$ combined as the Rossby number $R_W=U(fW_s)^{-1}$, where $W_s$ is the width at mid-length at the shelf-break isobath (alongshelf direction) (Fig. \ref{fig:sections}b).
\begin{table*}[t]
\caption{All runs in the Dynamical Experiment have a corresponding no-canyon run and constant vertical diffusivity as in the base case in Table \ref{tab:experiments_mix}. For all runs, parameters $N_0$, $f$ and $U$ were chosen to represent realistic oceanic conditions for canyons (within values in Table 1 in AH2010) while satisfying the dynamical restrictions imposed by AH2010 and HA2013. Only values changed from the base case (bold face entries in first row) are shown.}
\label{tab:experiments}
\begin{center}
\begin{tabular}{lcccccccc}
\hline \hline
Experiment & $N_0$ (s$^{-1}$) &	$f$ (s$^{-1}$) &	U (ms$^{-1}$)  & $Bu$ & $R_L$ & $R_W$ \\
\hline
 \textbf{base case}		& $\mathbf{5.5\times10^{-3}}$	& $\mathbf{9.66\times10^{-5}}$	& \textbf{0.36} & \textbf{0.40}	& \textbf{0.45}	& \textbf{0.31}	 \\
$\uparrow$ $N_0$		&$6.3\times10^{-3}$	&	&$0.38$	&$0.46$	&$0.47$	&$0.32$	\\ 
$\uparrow \uparrow$ $N_0$		&$7.4\times10^{-3}$	&	&$0.40$	&$0.54$	&$0.49$	&$0.33$	\\ 
$\downarrow$ $N_0$		&$5.0\times10^{-3}$	&	&$0.35$	&$0.37$	&$0.44$	&$0.30$	 \\
$\downarrow \downarrow N_0$		&$4.7\times10^{-3}$	&	&$0.35$	&$0.34$	&$0.43$	&$0.29$	 \\
$\Downarrow$ $N_0$		&$4.6\times10^{-3}$	&	&$0.35$	&$0.34$	&$0.43$	&$0.29$	 \\
$\uparrow f$		& &$1.00\times10^{-4}$	&$0.36$	&$0.39$	&$0.43$	&$0.29$	 \\
$\downarrow \downarrow$ $f$		&	&$7.68\times10^{-5}$	&$0.39$	&$0.51$	&$0.61$	&$0.41$	\\ 
$\downarrow f$		&	&$8.60\times10^{-5}$	&$0.38$	&$0.45$	&$0.53$	&$0.36$	 \\
$\Downarrow f$		&	&$6.40\times10^{-5}$	&$0.41$	&$0.61$	&$0.78$	&$0.53$	 \\
$\downarrow$ U		&	&	&$0.31$	&$0.40$	&$0.39$	&$0.26$	\\ 
$\downarrow \downarrow$ U		&	&	&$0.26$	&$0.40$	&$0.32$	&$0.22$	 \\
$\Downarrow$ U		&	&	&$0.14$	&$0.40$	&$0.18$	&$0.12$	 \\
$\Downarrow$ U, $\downarrow \downarrow$ $N_0$		&$4.6\times10^{-3}$	&	&$0.13$	&$0.34$	&$0.17$	&$0.11$	 \\
$\Downarrow$ U, $\uparrow \uparrow$ $N_0$		&$7.4\times10^{-3}$	&	&$0.15$	&$0.54$	&$0.19$	&$0.13$	 \\
$\Downarrow$ U, $\Downarrow$ $f$		&	&$7.00\times10^{-5}$	&$0.15$	&$0.56$	&$0.27$	&$0.18$	 \\
\hline
\end{tabular}
\end{center}
\end{table*}

\begin{table}[t]
\caption{All runs in the Mixing Experiment have the same dynamical parameters as the base case in Table \ref{tab:experiments}. All runs reported have a corresponding no-canyon run. Only values changed from the base case (bold face entries in first row) are shown. Values of $R_L$ and $R_W$ for these runs slightly vary from the base case values with $R_L$ between 0.42 and 0.44, and  $R_W$ between 0.28 and 0.30}
\label{tab:experiments_mix}
\begin{center}
\begin{tabular}{lccc}
\hline \hline
Experiment & $K_{bg}$ / m$^2$s$^{-1}$ & $K_{can}$ / m$^2$s$^{-1}$ & $\epsilon$ / m \\
\hline
\textbf{Base}	&$\mathbf{10^{-5}}$	&$\mathbf{10^{-5}}$	&\textbf{5} \\
$\uparrow K_{bg}$	&$10^{-4}$	&$10^{-4}$	&$5$ \\
$\uparrow \uparrow K_{bg}$	&$10^{-3}$	&$10^{-3}$	&$5$\\ 
$\Downarrow U \uparrow \uparrow K_{bg}$	&$10^{-3}$	&$10^{-3}$	&$5$ \\
$\Uparrow \Uparrow \uparrow K_{can}$	&	&$1.2\times10^{-2}$	&$5$ \\
$\Uparrow \Uparrow K_{can}$	&	&$10^{-2}$	&$5$ \\
$\Uparrow \Uparrow K_{can}$ $\epsilon 10$	&	&$10^{-2}$	&$10$ \\
$\Uparrow \Uparrow K_{can}$ $\epsilon 15$	&	&$10^{-2}$	&$15$ \\
$\Uparrow \Uparrow K_{can}$ $\epsilon 25$	&	&$10^{-2}$	&$25$ \\
$\Uparrow \Uparrow K_{can}$ $\epsilon 50$	&	&$10^{-2}$	&$50$ \\
$\Uparrow \Uparrow K_{can}$ $\epsilon 75$	&	&$10^{-2}$	&$75$ \\
$\Uparrow \Uparrow K_{can}$ $\epsilon 100$	&	&$10^{-2}$	&$100$ \\
$\Uparrow \uparrow \uparrow K_{can}$	&	&$8\times10^{-3}$	&$5$ \\
$\Uparrow \uparrow K_{can}$	&	&$5\times10^{-3}$	&$5$ \\
$\Uparrow \uparrow K_{can}$ $\epsilon 25$	&	&$5\times10^{-3}$ &$25$ \\
$\Uparrow \uparrow K_{can}$ $\epsilon 100$	&	&$5\times10^{-3}$ &$100$ \\
$\Uparrow K_{can}$	&	&$2.5\times10^{-3}$	&$5$ \\
$\uparrow \uparrow K_{can} $	&	&$10^{-3}$	&$5$ \\
$\uparrow \uparrow K_{can}$ $\epsilon 25$	&	&$10^{-3}$	&$25$ \\
$\uparrow \uparrow K_{can}$ $\epsilon 100$	&	&$10^{-3}$	&$100$ \\
$\uparrow K_{can}$	&	&$5\times10^{-4}$	&$5$ \\
\hline

\hline
\end{tabular}
\end{center}
\end{table}

\section{Results}
\label{sec:results}
\subsection{Description of the flow}
\label{sec:results_flow}
\begin{figure*}[th!]
\begin{center}
	\noindent\includegraphics[width=0.85\textwidth]{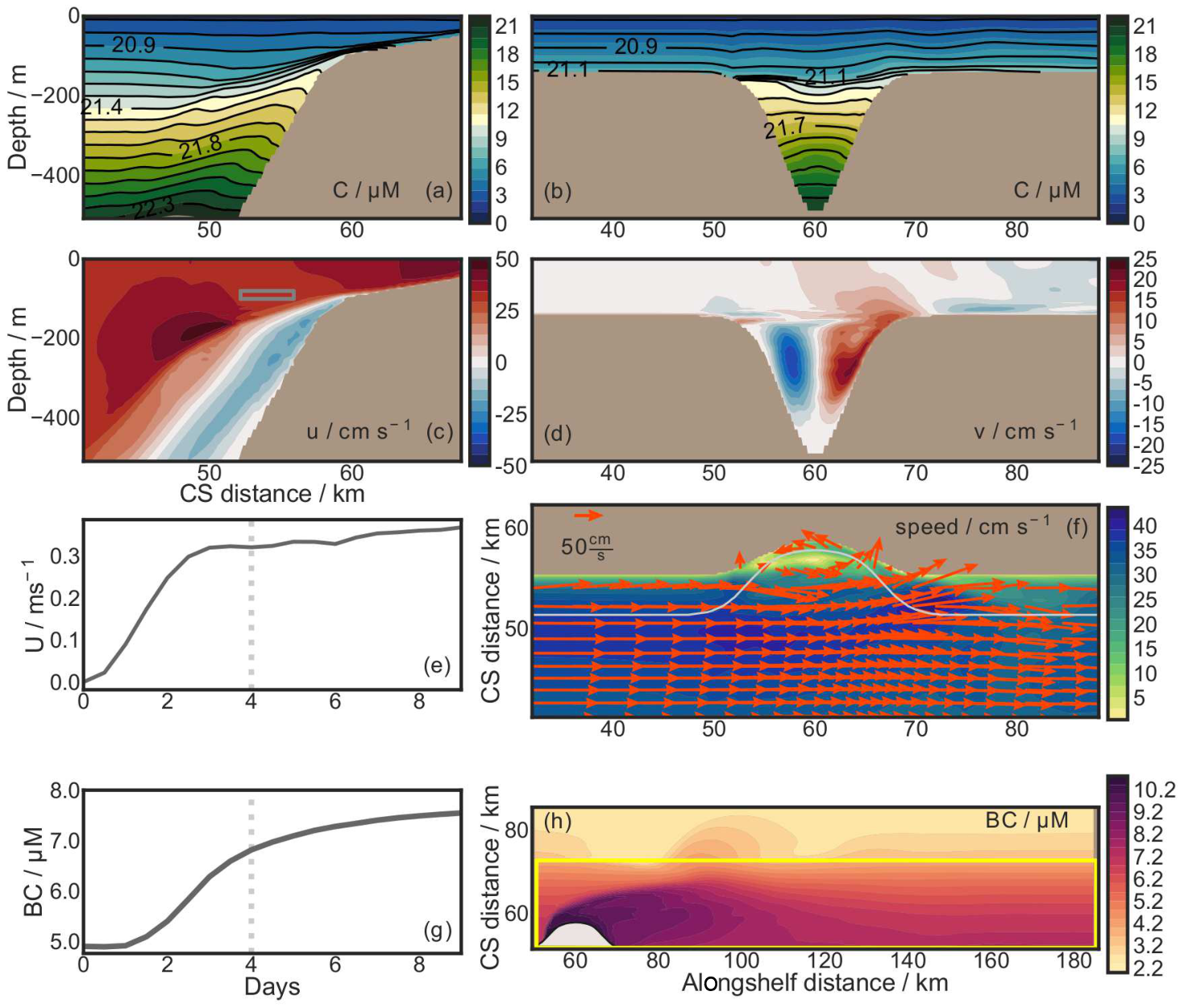}
	\caption{Main characteristics of the flow during the advective phase. Average day 3-5 contours of tracer concentration (color) and sigma-t (solid black lines, units of kgm$^{-3}$) at an alongshelf section close to canyon mouth (b) and along the canyon axis (a). (c) Along-shelf and (d) cross-shelf components of velocity. (e) Evolution of alongshelf component of incoming velocity $U$, calculated as the mean in the gray area delimited in (c). The dashed line marks the beginning of the advective phase. (f) Speed contours and velocity field at 127.5~m depth with shelf break in white. (h) Tracer concentration on the shelf bottom averaged over days 3-5. (g) Evolution of average bottom concentration on the downstream shelf, bounded by the yellow rectangle in (h). }
	\label{fig:flow}
	\end{center}
\end{figure*}

The body forcing generates an upwelling-favorable shelf current that slightly accelerates after the initial push (Fig. \ref{fig:flow}e). These conditions tilt the sea surface height down towards the coast. During spin up (days 0-3), the upwelling response is intense on the shelf and through the canyon. This time-dependent response is linear and thus, directly proportional to the forcing \citep{Allen1996} and will not be discussed further (time-dependent phase). We focus on the next stage, after day 4, when the current has been established, baroclinic adjustment has occurred, and advection dominates the dynamics in the canyon (advective stage). 

The highest alongshelf velocities can be found on the slope, at about 200 m (Fig. \ref{fig:flow}c), but the scale velocity for canyon upwelling is on the upstream-canyon shelf, between shelf break and canyon head, close to shelf bottom but above the bottom boundary current (Gray box, Fig. \ref{fig:flow}c). This constitutes the incoming velocity scale $U$ (Section \ref{sec:scaling}). In that area, $U$ stays between 0.35~ ms$^{-1}$ and 0.37 ms$^{-1}$ during the advective phase for the base case (Fig. \ref{fig:flow}e).  

Circulation over the canyon is cyclonic. An eddy forms near the canyon rim (Fig. \ref{fig:flow}f) and incoming flow deviates towards the head on the downstream side of the canyon and offshore on the downstream shelf. Within the canyon, below shelf-break depth, circulation is also cyclonic. Water comes into the canyon on the downstream side (positive v) and out on the upstream side (Fig. \ref{fig:flow}d). This circulation pattern is consistent with previous numerical investigations (\cite{Spurgin2014}, HA2013 and \cite{Dawe2010}), observations (\cite{Allen2001} and \cite{Hickey1997}) and laboratory experiments \citep{Mirshak2005}. 

Upwelling within the canyon is forced by an unbalanced horizontal pressure gradient between canyon head and canyon mouth \citep{Freeland1982}. In response, a balancing, baroclinic pressure gradient is generated by rising isopycnals towards the canyon head. The effect on the density field drives a similar response on the tracer concentration field (Fig. \ref{fig:flow}a). Near the canyon rim, pinching of isopycnals occurs on the upstream side (Fig. \ref{fig:flow}b). This region is associated with stronger cyclonic vorticity generated by incoming shelf water falling into the canyon, stretching the water column (Not shown). This well-known feature has been observed in Astoria Canyon \citep{Hickey1997} and numerically simulated (\emph{e.g.} HA2013 and \cite{Dawe2010}).

Most water upwells onto the shelf over the downstream side of the rim, near the canyon head. This upwelled water has higher tracer concentration than the water originally on shelf since the initial tracer profile increases with depth (Fig. \ref{fig:sections}f). As a result, a `pool'\, of water with higher tracer concentration than background values forms near shelf-bottom (Fig. \ref{fig:flow}h). This pool grows rapidly during the time-dependent phase, and more slowly during the advective phase (Animation S1 in SI). A similar feature was seen in a numerical study of canyon upwelling on the shelf of Washington, USA \citep{Connolly2014}. The average concentration near shelf bottom increases quickly during days 0 to 3 (by 1.5 $\mu$M) and more slowly during the next 6 days for the base case (Fig. \ref{fig:flow}g).

We isolate the canyon effect on the on-shelf tracer distribution by subtracting the corresponding no-canyon run. We look at the near-bottom tracer concentration anomaly (BC anomaly) defined as the concentration difference near the shelf bottom between the canyon and no-canyon case, normalized by the initial concentration near the bottom and expressed as a percentage. Contours of BC anomaly for the base case show a region of positive anomaly or higher tracer concentration relative to the no-canyon case downstream of the canyon (Animation S2 in SI). The vertical extent of the pool can be between 10 m and 30 m above the shelf bottom. The formation dynamics, extension and persistence of the pool will be characterized in future papers.

\subsection{Vertical gradient of density and tracer}
\label{sec:isopyc_isotracer}

\begin{figure*}[th!]
\begin{center}
	\noindent\includegraphics[width=0.8\textwidth]{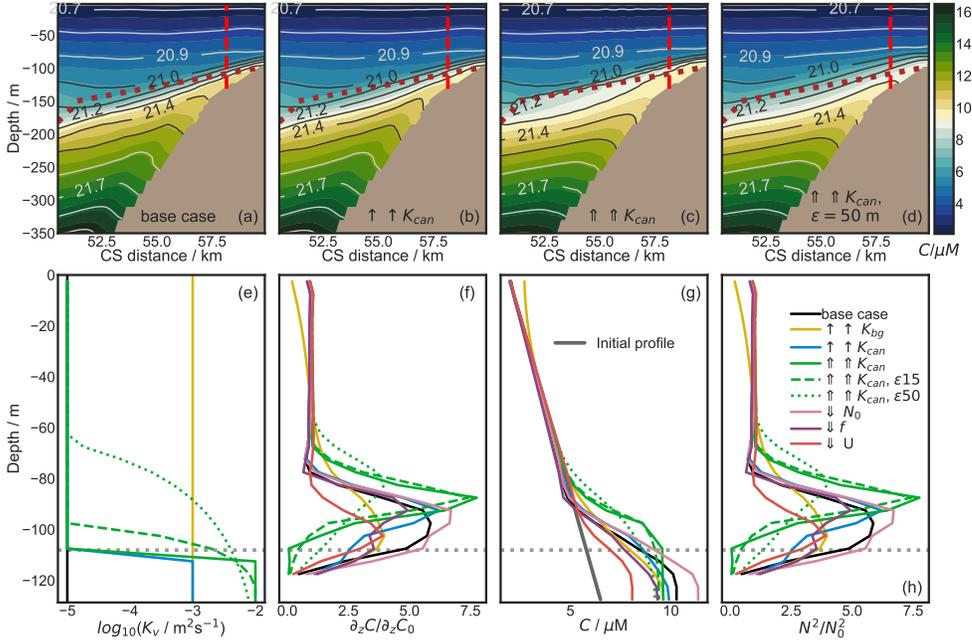}
	\caption{Concentration contours averaged over days 4 and 5 are plotted along canyon axis for the base case (a) and locally enhanced diffusivity cases with $K_{can}=10^{-3}$ m$^2$s$^{-1}$ (b), $K_{can}=10^{-2}$ m$^2$s$^{-1}$  (c) and $K_{can}=10^{-2}$ m$^2$s$^{-1}$, $\epsilon=50$ m (d). The dotted line indicates the location of the shelf downstream (rim depth). (e) Profiles of vertical diffusivity at a station in the canyon indicated by the dashed line in (a-d). (f, g, h) Vertical profiles of vertical tracer gradient divided by initial tracer gradient $(\partial_z C/\partial_z C_0)$, concentration $C$, and stratification divided by initial stratification $N^2/N^2_0$ taken on day 5 at same station as (e). Horizontal, dotted, grey lines correspond to the rim depth.}
	\label{fig:Tr_evolution}
	\end{center}
\end{figure*}

During an upwelling event, isopycnals and iso-concentration lines near the canyon rim are squeezed as they tilt up from mouth to head (Fig. \ref{fig:Tr_evolution}a). Close to the canyon head, on the downstream side where most upwelling occurs, stratification $N^2$ increases in time from the initial value $N_0^2$, with a maximum increase located close to but above rim depth at 108 m (Fig. \ref{fig:Tr_evolution}h). The maximum stratification increases quickly during the time-dependent phase of upwelling. After day 3, maximum stratification oscillates around the adjusted value; however, it slightly decreases for cases with enhanced $K_{can}$ and $K_{bg}$ because diffusivity weakens the density gradient with time. Maximum stratification above rim depth can be more than seven times higher than $N_0^2$. 

The tilting of isopycnals and thus, the increase in stratification near the head, is a baroclinic response to the unbalanced pressure gradient on the shelf. AH2010 showed that the pressure gradient along the canyon is $\rho_0 f U \mathcal{F}(Ro)$, where $\mathcal{F}(Ro)$ takes values between 0 and 1, which is consistent with our results: the maximum stratification increase is proportional to $U$ (compare pink to black line, Fig. \ref{fig:Tr_evolution}h) and $f$ (compare purple to black line, Fig. \ref{fig:Tr_evolution}h). They also show that the depth of the deepest isopycnal to upwell onto the shelf $Z$ is proportional to $N_{0}^{-1}$ (compare red to black line, Fig. \ref{fig:Tr_evolution}h). The deeper $Z$ is, the larger the tilting of isopycnals will be and so the larger the increase in stratification.

When diffusivity is locally enhanced, there is an additional effect on stratification. Within the canyon, enhanced diffusivity $K_{can}$ is acting on the density gradient, which was sharpened by the canyon-induced tilting of isopycnals, more rapidly than it is being diffused above the rim. So stratification near the rim but within the canyon is lower than it would be if the diffusivity profile was uniform, and stratification above the rim is higher than for the case with uniform diffusivity. The effect increases with $K_{can}$ (blue and solid green lines in Fig.\ref{fig:Tr_evolution}a, b and c) and it is maximum when the $K_v$ profile is a step (solid green line). Smoother $K_v$ profiles decrease the effect especially for $\epsilon$ larger than 25 m (dotted and dashed green lines).

Iso-concentration lines mimic isopycnals (Fig. \ref{fig:Tr_evolution}a-d, f). Vertical tracer gradients sharpen at rim depth as upwelling evolves, similar to stratification (Fig. \ref{fig:Tr_evolution}f). Compared to the base case, lower $N_0^2$ increases the sharpening effect on the tracer gradient (pink line vs. black line). 

Tracer concentration is relatively higher above rim depth with higher $K_{can}$ and lower below rim depth (all green and blue lines vs. black line above and below rim depth). This increased concentration is related to the gradient spike above rim depth (Fig. \ref{fig:Tr_evolution}g, green and blue lines).

\subsection{Cross-shelf transport of water and tracer}
\label{sec:CStrans}

To determine the pathways of water and tracers onto the shelf we calculate their cross-shelf (CS) and vertical transports. We define CS transport of water as the volume of water per unit time that flows across the vertical planes (CS1-CS6) that extend from the shelf-break in the no-canyon case to the surface (Fig. \ref{fig:sections}a and c); while vertical transports flow across the horizontal plane (LID) delimited by the shelf-break depth in the canyon case and the canyon walls (Fig. \ref{fig:sections}a and b).
\begin{figure*}[th!]
\begin{center}
	\noindent\includegraphics[width=0.85\textwidth]{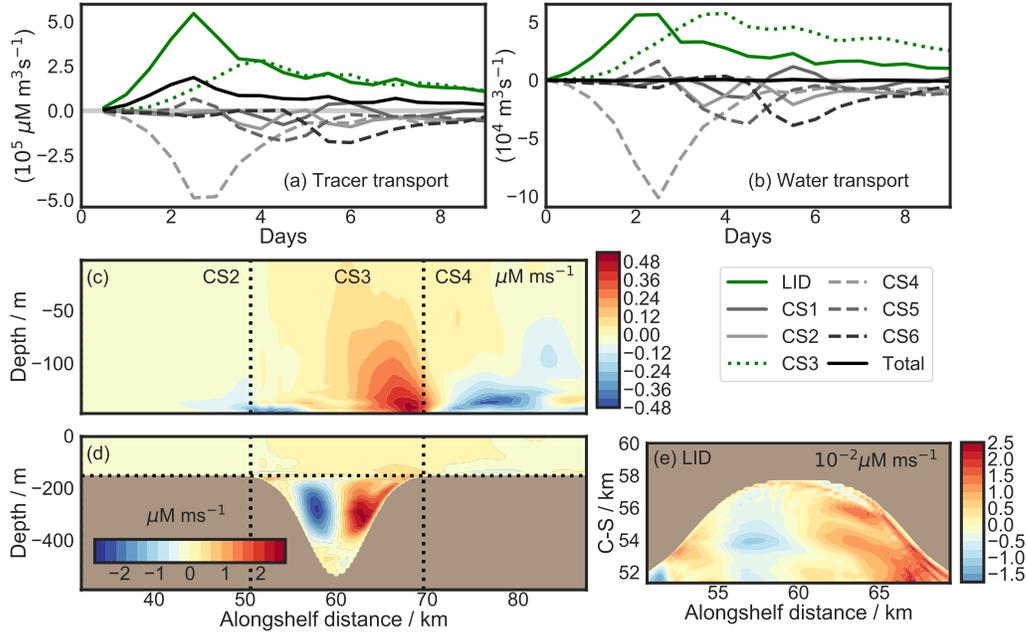}
	\caption{Base case (a) tracer and (b) water cross-shelf and vertical transport anomalies during the simulation through cross-sections defined in Figure \ref{fig:sections}. The horizontal, cross-shelf tracer transport anomaly through sections CS2, CS3 and CS4 and averaged over days 4 to 9 is plotted in (c). A full-depth version is plotted in (d) to include cross-shelf transport anomaly  through the canyon. Vertical tracer transport averaged over days 4 to 9 is plotted in (e).}
	\label{fig:CS_Transport}
	\end{center}
\end{figure*}

We define the net or total water and tracer transport onto the shelf (TWT and TTT, respectively) as the mean during the advective phase of the sum of the water and tracer transports through cross sections CS1 to CS6 and LID, and the vertical water transport (VWT) and tracer transport (VTT) onto the shelf as the mean transport through LID during the advective phase (days 4 to 9).

Tracer transport is divided into advective and diffusive contributions. The advective part is defined as $C\vec{u} \cdot \hat{n}A$, the contribution of the flow, where $\hat{n}A$ is the area vector normal to the cross-section. 
We compare the advective part to water transport. The flux and transport of tracers come directly from model diagnostics. 

The canyon effect in cross-shelf fluxes is the anomaly between canyon and no-canyon cases. Negative transports generally mean that either water or tracer are leaving the shelf; it is only near the shelf bottom, where shelf upwelling is onshore, that negative transports mean that transport for the no-canyon case is larger than in the canyon case. 

Patterns of cross-shelf transport anomaly of tracer and water are similar. Both, tracer (Fig. 4c) and water (not shown) anomaly flux is onto the shelf through CS3, close to the downstream side of the canyon mouth and through LID (vertical flux, Fig. 4e). Tracer and water transport anomalies flux off the shelf, downstream of the canyon, close to canyon mouth (CS4), and both transport anomalies are mainly offshore through CS1, CS2, CS5 and CS6. These agree with shelf-break upwelling suppression in the presence of a canyon. Deeper-than-shelf-break-depth water comes into the canyon through the downstream side and leaves through the upstream side, consistent with cyclonic circulation (Fig. \ref{fig:CS_Transport}d).


Positive transports through LID and CS3 indicate that tracer and water upwell onto the shelf throughout the simulation. The vertical upwelling response is maximum at the same time that the body forcing is maximum and then decreases to a steady value of 20\% of the maximum. Cross-shelf transport through CS3 reaches its maximum at day four and then decreases to a steady value of 40\% of that maximum (Fig. \ref{fig:CS_Transport}a, b). In contrast, transport anomaly through CS4, CS1, CS2, CS5 and CS6 is off-shore throughout the nine days. Off-shore transport at CS4 is the main balance to the onshore transports, especially during the time-dependent phase. Its response has similar timing as that of the vertical transport and it also decreases to a quasi-steady value after reaching its maximum on day 2.5. This off-shore transport is consistent with the off-shore steering of the flow described in section \ref{sec:results}\ref{sec:results_flow}. 

Overall, total tracer mass transport anomaly (TTT) is onto the shelf (Fig. \ref{fig:CS_Transport}a) and total water transport anomaly (TWT) is zero (Fig. \ref{fig:CS_Transport}b). During the advective phase there is a constant supply of tracer onto the shelf induced by the canyon (TTT). 
This result could change if the initial tracer profile was not linear or would reverse if it decreased with depth.

Changing dynamical parameters $R_W$ and $B_u$ changes the amount of transport relative to the base case, but qualitatively follows the same evolution through each section (Not shown). Higher (lower) $B_u$ and lower (higher) $R_W$ than in the base case decrease (increase) the amount of tracer transported onto shelf during the advective phase (Table \ref{tab:CSTrans}, column TTT). Enhanced $K_{can}$ increases the mean tracer transport onto the shelf (Table \ref{tab:CSTrans}, column TTT). This increase can be more than double when $K_{can}$ is two orders of magnitude larger than in the base case and triple with smoother profiles ($\epsilon >25$ m). 
\begin{table*}
\caption{Mean vertical (VTT), advective (VATT) and total (TTT) tracer transport anomalies through cross sections CS1-CS5 and LID as well as vertical water (VWT) and total  (TWT) water transport anomalies throughout the advective phase with corresponding standard deviations calculated as 12 hour variations for selected runs. Results for all runs are available in Table S1 in the supplementary material.}
\label{tab:CSTrans}
\begin{center}
\begin{tabular}{cccccc}
\hline \hline
Exp &	VTT   &	VATT &	TTT  &	VWT &	TWT  \\
        &	 $10^{5}$ $\mu$Mm$^3$s$^{-1}$ &	$10^{5}$ $\mu$Mm$^3$s$^{-1}$&	 $10^{4}$ $\mu$Mm$^3$s$^{-1}$&	 $10^{4}$ m$^3$s$^{-1}$&	 $10^{2}$ m$^3$s$^{-1}$ \\
\hline
base case	&1.6$\pm$0.29	&1.6$\pm$0.29	&0.46$\pm$0.13	&1.9$\pm$0.46	&-1.6$\pm$5.2	 \\
$\uparrow \uparrow$ $N_0$	&0.73$\pm$0.20	&0.73$\pm$0.20	&0.14$\pm$0.06	&0.91$\pm$0.29	&-5.5$\pm$4.7	 \\
$\downarrow \downarrow N_0$	&2.2$\pm$0.36	&2.2$\pm$0.36	&0.74$\pm$0.16	&2.5$\pm$0.58	&-1.4$\pm$4.0	 \\
$\uparrow f$	&1.7$\pm$0.32	&1.7$\pm$0.32	&0.49$\pm$0.13	&2.0$\pm$0.48	&-0.57$\pm$5.04	 \\
$\Downarrow f$	&0.92$\pm$0.16	&0.92$\pm$0.16	&0.21$\pm$0.09	&1.0$\pm$0.32	&-18.1$\pm$6.5	\\ 
$\Downarrow$ U	&0.43$\pm$0.05	&0.43$\pm$0.05	&0.12$\pm$0.01	&0.60$\pm$0.09	&4.3$\pm$2.2	 \\
$\Downarrow$ U, $\Downarrow$ $f$	&0.26$\pm$0.04	&0.26$\pm$0.04	&0.06$\pm$0.01	&0.37$\pm$0.07	&0.91$\pm$2.2\\
$\Uparrow \Uparrow K_{can}$, $\epsilon 25$	&2.5$\pm$0.18	&2.2$\pm$0.18	&1.2$\pm$0.06	&2.4$\pm$0.24	&4.6$\pm$1.1	\\ 
$\Uparrow \Uparrow K_{can}$, $\epsilon 50$	&2.8$\pm$0.15	&2.5$\pm$0.16	&1.3$\pm$0.06	&2.8$\pm$0.22	&3.0$\pm$1.3	\\ 
$\Uparrow \Uparrow K_{can}$, $\epsilon 100$	&3.0$\pm$0.15	&2.6$\pm$0.16	&1.3$\pm$0.08	&3.0$\pm$0.21	&1.8$\pm$1.3	\\ 
$\Uparrow \uparrow \uparrow K_{can}$	&2.2$\pm$0.25	&1.9$\pm$0.24	&1.0$\pm$0.10	&2.2$\pm$0.32	&9.5$\pm$2.2	\\ 
$\Uparrow \Uparrow K_{can}$	&2.0$\pm$0.25	&1.8$\pm$0.24	&0.97$\pm$0.11	&2.1$\pm$0.32	&9.6$\pm$1.8	 \\
\hline
\end{tabular}
\end{center}
\end{table*}

Upwelling through the canyon is well characterized by the vertical transport through LID. Vertical tracer transport (VTT) is dominated by advection over diffusion (VTT and the advective component VATT are equal to 2 significant figures for runs in dynamical experiments, not all shown). Nonetheless, the advective component (VATT) is modified by enhanced vertical diffusivity through modifications to the density field. Larger diffusivities weaken the density gradients near the rim which allows more water to upwell onto the shelf. During the advective phase, VATT tends to increase when diffusivity is enhanced and can be as much as 25\% larger than in the base case (Table \ref{tab:CSTrans}). VTT can be higher by 25\% to 37\% for the largest two $K_{can}$ used (Table \ref{tab:CSTrans}) and can almost double for smoother $K_v$ profiles ($\epsilon=100$ m). The effect of enhanced $K_{can}$ amplifies throughout the simulation depending on the magnitude of $K_{can}$ and the gradient.

\subsection{Upwelling flux and upwelled tracer mass}
\label{sec:HCW}
Upwelled water on the shelf has been estimated previously by finding water originally below shelf-break depth based on its salinity (HA2013). We take the same approach but use the tracer concentration at shelf-break depth as the criterion to find water on shelf that was originally below shelf-break depth. For this we use the low diffusivity tracer described in Section \ref{sec:Methodology} . 
\begin{figure*}[th!]
\begin{center}
	\noindent\includegraphics[width=0.75\textwidth]{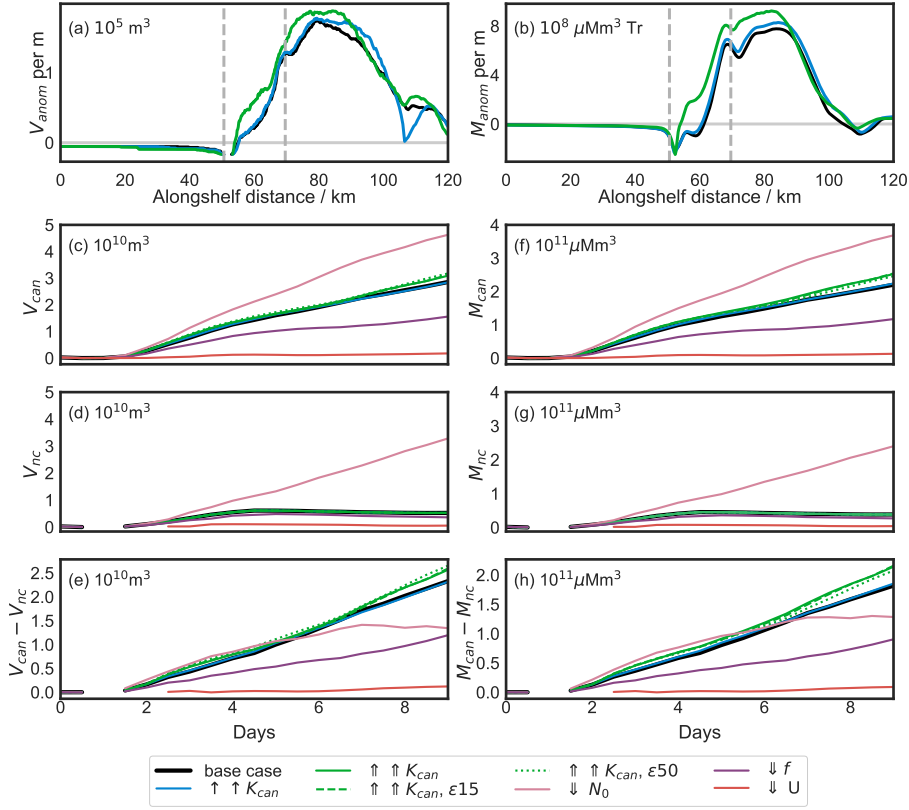}
	\caption{ (a-b) Vertically-integrated upwelled water volume and tracer at day 3.5 alongshelf. Dashed lines show the position of the canyon. (c, d, e) Volume of water on the shelf with concentration values initially below shelf-break depth. (f, g, h) Upwelled tracer mass on shelf. (c) and (f) canyon cases, (d) and (g) no-canyon cases, and  (e) and (h) the difference between these. The boundaries of the shelf box are the wall that goes from shelf-break to surface in the no-canyon case, alongshelf wall at northern boundary and cross-shelf walls at east and west boundaries.  }
	\label{fig:HCW_TrMass_CanNoC}
	\end{center}
\end{figure*}
We define the volume of water upwelled onto the shelf through the canyon ($V_{anom}$) at day $t$ as the difference between the volume of upwelled water, that is water with $C>C_s=7.2$ $\mu$M, where $C_s$ is the initial concentration at shelf-break depth, on shelf at $t=t$ in the canyon case ($V_{can}$) compared to the no canyon case ($V_{nc}$):
 
\begin{equation}
V_{anom}(t) =  V_{can}-V_{nc} = \sum_{can}\Delta V  - \sum_{nc}\Delta V \, \, \textrm{where } C>C_{s}
\label{eq:Vanom}
\end{equation}
  
where $\Delta V$ is the volume of the cell with concentration higher than that at shelf-break depth ($C_{s}$) and the sum is over all cells on the shelf that satisfy this criterion for the bathymetry with a canyon (sum over $can$) and without a canyon (sum over $nc$). The shelf volume constitutes all the cells between the shelf break and the coast, and between the shelf bottom and the surface. 
Similarly, the tracer mass upwelled by the canyon ($M_{anom}$) is defined as the tracer mass contained within the upwelled water $V_{anom}$
  
\begin{equation}
M_{anom}(t) =  M_{can}-M_{nc} = \sum_{can}C\Delta V  - \sum_{nc}C\Delta V \, \, \textrm{where } C>C_{s}
\label{eq:Manom}
\end{equation}
  
 where $C$ is the concentration at the cell, $\Delta V$ is the volume of the cell and the sum is over all cells with upwelled water. 

Water upwells onto the shelf on the downstream side of the canyon rim. The upwelled-water volume anomaly, $V_{anom}$ (\ref{eq:Vanom}) alongshelf (integrated in the cross-shore direction) at day 3.5 is concentrated on the shelf, on the downstream side of the canyon rim (Fig. \ref{fig:HCW_TrMass_CanNoC}a). Water continues upwelling through the canyon and, at the same time, the bulge of upwelled water is advected downstream. On the upstream shelf, shelf-break upwelling is suppressed as water is redirected to upwell through the canyon as indicated by negative values of upwelled water volume anomaly (Fig. \ref{fig:HCW_TrMass_CanNoC}a). The upwelled tracer mass anomaly, $M_{anom}$ as in (\ref{eq:Manom}) follows a similar pattern alongshelf (Fig. \ref{fig:HCW_TrMass_CanNoC}b). 

The upwelled water volume ($V_{can}$) through the canyon is larger than that upwelled on a straight shelf and increases throughout the simulation. In the canyon case, water upwelling is dominantly canyon induced; at day 9, it accounts for between 24\% to 89\% of $V_{can}$ throughout the runs and between 25 to 90\% of upwelled tracer mass ($M_{can}$) (\ref{tab:HCW_TrMass}), except for the lowest $U$ case with enhanced background diffusivity, where canyon-induced upwelling accounts for 0.8\%.  
\begin{table*}
\caption{Mean water and tracer upwelling fluxes ($\Phi$ (\ref{eq:phi}) and $\Phi_{Tr}$ (\ref{eq:phiTr})) for selected runs during the advective phase, reported with 12 hour standard deviations. All other quantities are evaluated at day 9:  Volume of upwelled water ($V_{can}$), upwelled tracer ($M_{can}$) for the canyon case and  fractional canyon contributions to these quantities calculated as the canyon case minus the no-canyon case divided by the canyon case, and total tracer mass anomaly on shelf ($\mathcal{M}$-$\mathcal{M}_{nc}$ (\ref{eq:Mcal})) in kg of NO$^-{_3}$. Results for all runs are available in Table S2 in the supplementary material.}
\label{tab:HCW_TrMass}
\footnotesize
\begin{tabular}{cccccccc}
\hline \hline
Exp &	
\multicolumn{1}{p{1.5cm}}{\centering$\Phi$ \\ $(10^4$ m$^3$s$^{-1})$} &
\multicolumn{1}{p{1.9cm}}{\centering$\Phi_{Tr}$ \\ $(10^5$ $\mu$Mm$^3$s$^{-1})$} &
\multicolumn{1}{p{1.2cm}}{\centering $V_{can}$  \\ $(10^{10}$ m$^3)$} &
\multicolumn{1}{p{1.4cm}}{\centering $(V_{can}-V_{nc})$  $V_{can}^{-1}$ (\%)} &
\multicolumn{1}{p{1.6cm}}{\centering $M_{can}$ \\ $(10^{11}$ $\mu$Mm$^3)$} &
\multicolumn{1}{p{1.5cm}}{\centering $(M_{can}-M_{nc})$  $M_{can}^{-1}$ (\%)} &
\multicolumn{1}{p{1.6cm}}{\centering $\mathcal{M}-\mathcal{M}_{nc}$ \\ ($10^6$ kg NO$^-_3$)} \\
\hline
base case	&3.85$\pm$0.60	&2.76$\pm$ 0.26	&2.86	&81.61		&2.20		&82.57	&1.96 \\
$\uparrow \uparrow$ $N_0$	&1.32$\pm$0.55	&1.11$\pm$ 0.44	&1.10	&77.74		&0.82		&77.91		&0.62\\
$\downarrow \downarrow N_0$	&6.34$\pm$0.92	&4.84$\pm$ 0.72	&4.35	&30.84		&3.45		&36.36		&3.25\\
$\uparrow f$	&4.03$\pm$0.58	&2.95$\pm$ 0.36	&2.96	&73.08		&2.30		&74.70		&2.06\\
$\Downarrow f$	&1.83$\pm$0.88	&1.03$\pm$ 0.42	&1.56	&76.77		&1.17		&77.09		&1.03\\
$\Downarrow$ U	&0.14$\pm$0.23	&0.15$\pm$ 0.07	&0.18	&69.05		&0.13		&69.53		&0.31\\
$\uparrow$ $K_{bg}$	&3.70$\pm$0.73	&2.29$\pm$ 0.24	&2.80	&87.26		&2.05		&87.20		&2.02\\
$\Uparrow \Uparrow K_{can}$, $\epsilon 25$	&4.12$\pm$0.71	&3.43$\pm$ 0.50	&3.14	&83.29		&2.52		&84.76		&4.24\\
$\Uparrow \Uparrow K_{can}$, $\epsilon 50$	&4.21$\pm$0.71	&3.29$\pm$ 0.55	&3.18	&83.48		&2.46		&84.38		&4.53\\
$\Uparrow \Uparrow K_{can}$, $\epsilon 100$	&4.51$\pm$0.64	&3.40$\pm$ 0.46	&3.36	&84.37		&2.52		&84.76		&4.70\\
$\Uparrow \uparrow \uparrow K_{can}$	&4.08$\pm$0.54	&3.35$\pm$ 0.33	&3.08	&82.94		&2.51		&84.70		&3.48\\
$\Uparrow \Uparrow \uparrow K_{can}$	&4.17$\pm$0.65	&3.51$\pm$ 0.39	&3.13	&83.22		&2.55		&84.95	&3.27\\
$\Uparrow \Uparrow K_{can}$	&4.09$\pm$0.58	&3.39$\pm$ 0.37	&3.09	&83.02		&2.53		&84.79	&3.38\\
\hline
\end{tabular}
\end{table*}
We calculate the upwelling flux $\Phi$ as the mean of the daily flux of $V_{can}$ during the advective phase (between day 4 and 9),
 
\begin{equation}
\Phi=\big \langle \frac{\partial}{\partial t}(V_{can}) \big \rangle.
\label{eq:phi}
\end{equation}
  
 We considered the full upwelled volume of water for the canyon case following the metric defined by HA2013 since we will compare our results to their scaling estimate and then use this estimate for our scaling of tracer upwelling flux (Sec. \ref{sec:scaling}). Consequently, we define the upwelled tracer mass flux $\Phi_{Tr}$ as the mean daily flux of $M_{can}$ between day 4 and 9,
 
\begin{equation}
\Phi_{Tr}=\big \langle \frac{\partial}{\partial t}(M_{can}) \big \rangle.
\label{eq:phiTr}
\end{equation}
  
\begin{figure}[th!]
\begin{center}
	\noindent\includegraphics[width=0.5\textwidth]{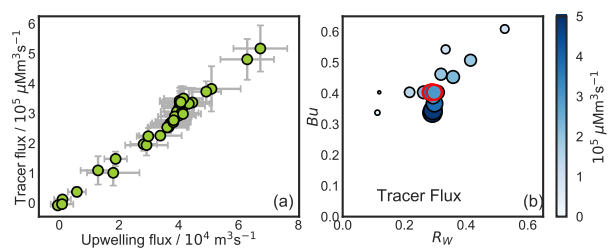}
	\caption{(a) Comparison between the mean flux of water and the mean flux of tracer upwelled  through the canyon during the advective phase of upwelling for all runs. Error bars correspond to standard deviations.  (b) Upwelled tracer flux increases (darker, larger markers) with increasing Rossby number $R_W$  and decreasing Burger number $Bu$. The size and color of the markers are proportional to the tracer flux, and the red-edged markers correspond to runs with locally-enhanced $K_{can}$. Locally-enhanced diffusivity weakens the stratification below rim depth and allows more tracer to upwell. }
	\label{fig:fluxes}
	\end{center}
\end{figure}

The upwelling tracer flux is directly proportional to the water upwelling flux, with small deviations when the $K_v$ profile is a step (Fig. \ref{fig:fluxes}a). Water and tracer upwelling fluxes (Table \ref{tab:HCW_TrMass}, columns 2 and 3, respectively) are inversely proportional to $Bu$ (for fixed $R_W$) and directly proportional to $R_W$ (for fixed $Bu$). This dependence of the upwelling flux of water with $Bu$ and $R_W$ is consistent with findings by AH2010 and HA2013; the same dependence of the tracer flux on $Bu$ and $R_W$ shows that the upwelling of tracers is dominated by advection (Fig. \ref{fig:fluxes}b). Cases with smaller (larger) $f$ and thus, simultaneously higher (lower) $R_o$ and $Bu$, have smaller (larger) upwelling fluxes. This is consistent with the relatively high values of $R_W$ that we are exploring.

Locally-enhanced diffusivity moderately increases the tracer upwelling flux and the water upwelling flux which increase by 27\% and 19\%, respectively compared to the base case for the highest $K_{can}$. Moreover, high $K_{can}$ combined with a smooth $K_v$ profile (e.g. $\epsilon=100$ m) increases the upwelling flux of water by 26\%. The results suggest that larger values of $\epsilon$ increase the water upwelling flux but the average increase is smaller than the standard deviation and so it cannot be confirmed, except by comparing the extreme cases, $\epsilon=5$ m and $\epsilon = 100$ m. Enhanced background diffusivity decreases the tracer mass upwelling flux as much as 61\% for the highest $K_{bg}$ although these cases are not physically relevant. 

We calculate the total amount of tracer mass on shelf at a given time ($\mathcal{M}(t)$) by integrating the volume of each cell on the shelf multiplied by its tracer concentration: 
 
\begin{equation}
\mathcal{M}(t)=\sum_{shelf}C \Delta V,
\label{eq:Mcal}
\end{equation}
  
where $\Delta V$ is the volume of a cell on the shelf and $C$ its concentration. This includes cells from the bottom of the shelf all the way to the surface and from shelf break to the coast. The total volume of the shelf is $6.1\times10^{11}$ m$^3$ and the volume of the canyon is approximately $6.8\times10^{9}$ m$^3$. So, the canyon represents about 1\% of the total volume of the shelf. $\mathcal{M}(t)$ reflects all processes and exchanges of mass at any depth and from any kind of water; it is the total inventory of tracer on shelf.

Given that the tracer we added had an initial linear nitrate profile, the difference in the total on-shelf nitrate inventory $\mathcal{M}$ as in (\ref{eq:Mcal}) between the canyon case and the straight shelf case can be between 0.3-4.7$\times10^6$ kg NO$^-_3$ after 9 days of upwelling (Table \ref{tab:HCW_TrMass}, last column). If we consider a two month upwelling period, our estimate for one canyon is 0.2-3.1$\times10^7$ kg NO$^-_3$. \cite{Connolly2014} numerically estimated the nitrate input of two canyons in the Washington Shelf during June and July to be between 1-2$\times10^7$ kg NO$^-_3$, which is consistent with our estimate.

\section{Scaling analysis }
\label{sec:scaling}
There are two main processes acting to transport tracer onto the shelf: mixing and advection. The mixing contribution, represented by locally-enhanced diffusivity within the canyon, has been described in the results and is scaled in this section, while the advective part is driven by the upwelling dynamics described and scaled by AH2010 and HA2013. Additionally, we found that enhanced mixing within the canyon can have an effect on advection through modifications to the density field near the canyon head and we include a correction for it. 

The scaling by HA2010 and HA2013 starts from the shallow-water equations for an inviscid, steady, Boussinesq flow. They characterize the tendency of the flow to follow the bathymetry to determine the strength of upwelling through the canyon. Then, they calculate the effective, unbalanced pressure gradient within the canyon that is responsible for raising the isopycnals. Next, they calculate the resulting density deformation, from which they identify the deepest isopycnal that upwells onto the shelf. The depth of this isopycnal is $H_h+Z$ where $H_h$ is the canyon-head depth and $Z$ is called the depth of upwelling:
 
\begin{equation}
Z ={\bigg(\frac{fUL\mathcal{F}}{N_0^2}\bigg)}^{1/2} 
\label{eq:Z}
\end{equation}
  
where $\mathcal{F}=R_o/(0.9+R_o)$  is the function that characterizes the tendency of the flow to cross the canyon and $R_o=U/f\mathcal{R}$ is a Rossby number that uses the upstream radius of curvature $\mathcal{R}$ as a length scale. Other useful estimates from this analysis are the horizontal $U_*$ and vertical $\Omega$ components of velocity of the upwelling current, given by 
 
\begin{align}
U_* &\approx  U\mathcal{F}  \label{eq:Ustar} \\
\Omega& = \frac{U_*Z}{L}\label{eq:Omega}
\end{align}
  
We use $U_*$ and $\Omega$ to find the relative importance of the terms in the advection-diffusion equation. Starting from the scales $Z$, $U_*$ and $\Omega$, AH2010 carry on to estimate the upwelling flux $\Phi=U_*W_{m}Z$ by arguing that the flux of upwelling is the flux coming into the canyon at the mouth (width $W_{m}$), over a depth $Z$ at speed $U_*$. They use observations and results from numerical and physical models to find the coefficients of the scaled quantities.

There were two criteria that guided our choice of the dynamical parameter space: To have realistic values of $U$, $N_0$ and $f$ in the context of shelf regions, and to satisfy the restrictions imposed by AH2010 and HA2013. There are 9 restrictions that apply to the scaling estimates for a canyon (AH2010 section 2.5, HA2013). In summary, the scaling requires the flow to be uniform over the length of the canyon, $L$,  and relatively weak ($\mathcal{F}R_W<0.2$). The stratification to be nearly uniform near canyon rim. The shelf break to be shallow enough that isopycnals over the canyon feel the canyon close to the surface, so that the effective depth over the canyon is the shelf-break depth $H_s$ ($B_s<2$ where $B_s=NH_s/fL$). The continental shelf must be sloped so that the onshore bottom boundary layer (BBL) flow is shut down. The canyon walls must be steep (so that BBL flows are quickly arrested), the canyon much deeper than the depth of upwelling $Z$, and the canyon width should be narrower than 2 Rossby radii. The scaling is general enough that it has been successfully compared to observations in six canyons, three laboratory models and a recent field study in Whittard Canyon \citep{Porter2016}.

\subsection{Advection-diffusion equation in natural coordinates}
\begin{figure}[th!]
\begin{center}
	\noindent\includegraphics[]{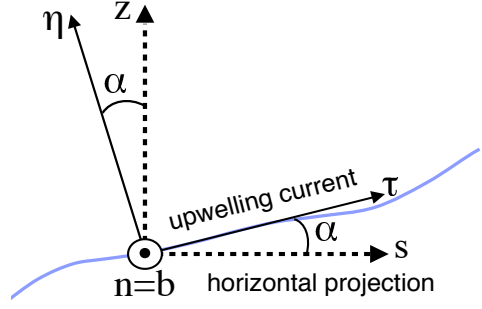}
	\caption{The coordinate system ($\tau$, $\eta$, $b$) corresponds to the trihedron that moves along the upwelling current (blue line) and ($s$, $z$, $n$) corresponds its horizontal projection (natural coordinate system).  So $\vec{s}$ is the projection of $\vec{\tau}$ in the $x-y$ plane (horizontal) and $z$ is the usual vertical coordinate ($s$, $z$, $n$). In our derivation, we assume that the coordinates $n$ and $b$ are the same and moreover, they lie on the isopycnal plane. This means that the difference between the coordinate systems is a single rotation $\alpha$ around the $n/b$ axis.}
	\label{fig:SysRef}
	\end{center}
\end{figure}

Let ($\hat{\tau}$, $\hat{\eta}$, $\hat{b}$) be a flow-following coordinate system that describes the motion of a trihedron along the curve given by the upwelling current. The unitary trihedron is defined by $\hat{\tau}$, the vector tangent to the upwelling current; $\hat{\eta}$ the normal vector to the tangent in the same vertical plane and pointing upwards, and $\hat{b}=\hat{\tau} \times \hat{\eta}$, the vector normal to the plane defined by $\hat{\tau}$ and $\hat{\eta}$ (Figure \ref{fig:SysRef}).

Let us consider the deepest streamline that upwells within a submarine canyon and assume that the isopycnal plane is associated with the $\tau-b$ plane, since the canyon-induced upwelling flow is favoured along isopycnals, and the diapycnal direction with $\hat{\eta}$.

The equation describing the change in concentration of a passive tracer $C$, is written in natural coordinates ($s$, $z$, $n$) \citep{Holton1992} as
 \begin{equation}
\frac{\partial{C}}{\partial{t}}+u\frac{\partial{C}}{\partial{s}}+w\frac{\partial{C}}{\partial{z}}=\nabla \cdot \mathcal{K} {\nabla}C,
\label{eq:adv-diff_snz}
\end{equation}
  
where $u$ is the horizontal velocity, $w$ is the vertical component of velocity, and $\cal{K}$ is an order 2 diffusivity tensor. However, the simplest representation of $\cal{K}$ is in isopycnal coordinates:
 
\[\mathcal{K}=
\begin{pmatrix}
  K_{I} & 0 & 0 \\
    0 & K_{I} & 0 \\
  0 & 0 & K_{D}
 \end{pmatrix}\]
  
where $K_{I}$ is the diffusion coefficient along isopycnals and $K_{D}$ is the diffusion coefficient in the diapycnal direction. To use this, we can express the rhs of (\ref{eq:adv-diff_snz}) in the coordinates ($\tau$, $\eta$, $b$), associated with the isopycnal-diapycnal directions as 
 
 \begin{equation}
\frac{\partial{C}}{\partial{t}}+u\frac{\partial{C}}{\partial{s}}+w\frac{\partial{C}}{\partial{z}}=K_{I} {\nabla}^{2}_{\tau,b}C+K_{D}\frac{\partial^2{C}}{\partial{\eta^2}}.
\label{eq:adv-diff_tau_eta_b}
\end{equation}
  
 Expressing the rhs of (\ref{eq:adv-diff_tau_eta_b}) in terms of ($s$,$z$,$n$) (Appendix A) we arrive at an equation in terms of isopycnal and diapycnal diffusivities, and the  upwelling current velocity components in natural coordinates:
  
 \begin{align}
  \frac{\partial{C}}{\partial{t}}+u\frac{\partial{C}}{\partial{s}}+w\frac{\partial{C}}{\partial{z}}&\approx K_{I}\left(\frac{\partial^2C}{\partial{s^2}}+\frac{\partial^2C}{\partial{n^2}}+2\frac{\partial^2C}{\partial{z}\partial{s}}\frac{\partial{z}}{\partial{\tau}}\right) \nonumber\\
  &+ K_{D}\left(\frac{\partial^2C}{\partial{z^2}}+2\frac{\partial^2C}{\partial{z}\partial{s}}\frac{\partial{s}}{\partial{\eta}}\right). 
  \label{eq:adv-diff_natural}
  \end{align}

\subsection{Relevant parameter space} 
The relevant dynamical variables in (\ref{eq:adv-diff_natural}) are the horizontal and vertical velocities $u$ and $w$, scaled by the horizontal upwelling velocity $U_*$ and vertical upwelling velocity $\Omega$, respectively; and the isopycnal and diapycnal diffusivity coefficients $K_{I}$ and $K_{D}$. Additional parameters are the scales for the horizontal and vertical concentration gradients $\delta_{h}C$ and $\delta_{v}C$; and scales for the horizontal and vertical curvatures of the concentration $\delta^2_{h}C$ and $\delta^2_{v}C$, respectively. The curvatures of the concentration are the second derivative of the concentration profile with respect to depth ($\partial^2{C}/\partial{z^2}$), and with respect to the cross-shelf direction $\partial^2{C}/\partial{s^2}$, within the canyon. 
 A horizontal length scale is given by the canyon length, $L$ and a vertical length scale by the depth of upwelling, $Z$. 

In total, there are 10 parameters ($U_*$, $\Omega$, $L$, $Z$, $\delta_{h}C$, $\delta_{v}C$, $\delta^2_{h}C$, $\delta^2_{v}C$, $K_{I}$ and $K_{D}$ ) with four dimensions:  horizontal length, vertical length, time and concentration. We differentiate between horizontal and vertical lengths because we are assuming that the flow is hydrostatic and thus, vertical and horizontal processes are decoupled. According to the Buckingham-$\Pi$ theorem \citep{Kundu2004} there are six non-dimensional groups that dynamically represent the system (Table \ref{tab:NonDim}).

In terms of these non-dimensional numbers, the advection-diffusion equation (\ref{eq:adv-diff_natural}) for the steady state can be expressed in non-dimensional form as
 
 \begin{align}
 \frac{KPe_h}{\Gamma}u'{\bigg(\frac{\partial C}{\partial s}\bigg)}'+
 Pe_vw^{\prime}{\bigg(\frac{\partial C}{\partial z}\bigg)}^{\prime} & =
 K\tau_h{\bigg[\bigg(\frac{\partial^2 C}{\partial s^2}\bigg)' + 
              2\bigg( \frac{\partial^2 C}{\partial z \partial s}\frac{\partial z }{\partial \tau}\bigg)'\bigg]} \nonumber \\
&  - \tau_v{\bigg[\bigg(\frac{\partial^2 C}{\partial z^2}\bigg)' + 2\bigg( \frac{\partial^2 C}{\partial z \partial s}\frac{\partial s }{\partial \eta }\bigg)'\bigg]},
\label{eq:adv-diff_non_dim}
\end{align}
  
where the primed variables are non-dimensional, \emph{e.g.} $u'=u(U_*)^{-1}$, $(\partial C/ \partial s)'=(\partial C/ \partial s)(\delta_hC)^{-1}$, etc. 

We estimate the scales $U_*$, $Z$ and $\Omega$, given by (\ref{eq:Z}), (\ref{eq:Ustar}) and (\ref{eq:Omega}), respectively, using as a test case Barkley Canyon. The relative importance of each parameter can be drawn from the values of these non-dimensional quantities (Table \ref{tab:NonDim}). 

Horizontal advection will dominate over isopycnal diffusivity ($Pe_h>>1$) and so we did not include it in the parameter space of our experiments. On the other hand, vertical advection and vertical diffusivity are both relevant for this flow ($Pe_v \approx O(1)$). Finally, the effect of vertical diffusivity is locally larger than that of isopycnal diffusivity ($K<<1$). 

Non-dimensional numbers $\Gamma$, $\tau_h$ and $\tau_v$ represent the competition between geometric characteristics of the initial vertical and horizontal tracer profiles. The role of these parameters and their implications will be discussed in future studies.  
\begin{table*}
\caption{Non-dimensional groups constructed for the tracer scaling. To calculate these scales, we took geometric parameters reported by \cite{Allen2001} for Barkley Canyon ($L$=6400~m, $\cal{R}$=5000~m , $W_{m}$=13000~m), stratification and incoming velocity values reported by \citet{Allen2010} ($N_0=10^{-3}$~s$^{-1}$, $U=0.1$~ms$^{-1}$). Although not measured for Barkley Canyon, we used the diapycnal diffusivity ($K_D=3.90\times10^{-3}$~m$^2$s$^{-1}$ \citep{Gregg2011} and isopycnal diffusivity $K_I=2$~m$^2$s$^{-1}$ \citep{Ledwell1998}. }
\label{tab:NonDim}
\begin{center}
\begin{tabular}{  c c  c c }
\hline \hline
Symbol & Definition & Description & Barkley Canyon estimate \\
\hline
& & & \\
 $Pe_h$       & \Large{ $\frac{LU_*}{ K_{I}}   $  }              & Horizontal Peclet number & $2.1 \times 10^2 $    \\      
& & &\\
 $Pe_v$       & \Large{$\frac{Z \Omega}{ K_{D}}$ } & Vertical Peclet number & $1.2$ \\       
& & &\\
 $K$            & \Large{$\frac{Z^2}{L^2}\frac{K_{I}}{K_{D}}$   }      & Diffusivity ratio & $5.9 \times 10^{-3} $  \\
& & &\\
 $\Gamma$ &  \Large{$\frac{Z}{L} \frac{\delta_{v}C}{\delta_{h}C} $ }        & Gradient ratio & Tracer dependent \\
& & &\\
 $\tau_h$                 & \Large{$\frac{ L \delta^2_{h}C}{\Gamma  \delta_{h}C}$} & Horizontal curvature to gradient ratio & Tracer dependent \\
& & &\\
 $\tau_v$                  & \Large{$-\frac{Z \delta^2_{v}C}{ \delta_{v}C}$} & Vertical curvature to gradient ratio & Tracer dependent     \\
& & &\\
\hline
\end{tabular}
\end{center}
\end{table*}

\subsection{Stratification and tracer gradient evolution}
\label{sec:tracer_gradient}
In our system, the evolution of isopycnals during canyon-induced upwelling is very similar to that of tracer iso-concentration lines, as shown in section \ref{sec:results}\ref{sec:isopyc_isotracer}; thus, vertical tracer gradient and stratification evolve similarly. 
\begin{figure*}[th!]
\begin{center}
	\noindent\includegraphics[width=0.7\textwidth]{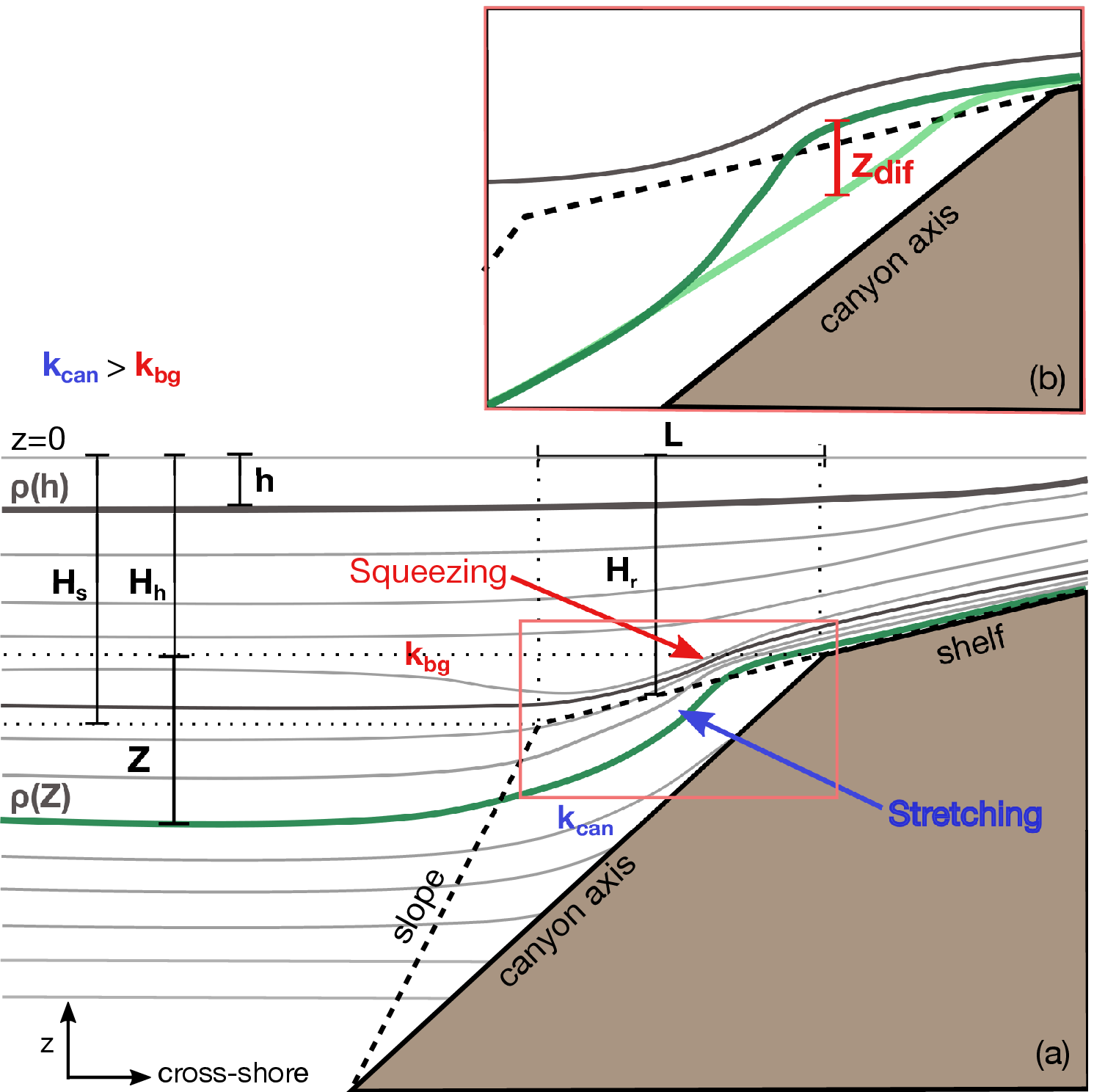}
	\caption{(a) Isopycnals (gray lines) tilt towards the canyon head during a canyon-induced upwelling event. This tilt is proportional to the upwelling depth $Z$, defined as the displacement of the deepest isopycnal to upwell onto the shelf (heavy, green line). Locally enhanced vertical diffusivity $K_{can}$ compared to the background value $K_{bg}$ further squeezes isopycnals above rim depth (canyon rim represented by the dashed line) and in turn, further stretches isopycnals below rim depth. The squeezing effect is proportional to the characteristic length $Z_{dif}$ (\ref{eq:Zdif}). (b) Zoom in of the red square keeping only two isopycnals: the heavy, dark green line is the deepest isopycnal that upwells onto the shelf and the grey one is a reference isopycnal. The light green line represents the deepest isopycnal that upwells when diffusivity is homogeneous everywhere (base case). The extra displacement of this isopycnal when diffusivity is locally-enhanced is the scale $Z_{dif}$.}
	\label{fig:scaling_schematics}
	\end{center}
\end{figure*}

 During the advective phase of upwelling, isopycnals will squeeze near the head of the canyon, increasing the stratification with respect to the initial value (Fig. \ref{fig:scaling_schematics}). Near the downstream side of the canyon rim the amplification of stratification (the ``squeezing'')  can be expressed as
 
\begin{equation}
S = \frac{N^2}{N^2_0},
\end{equation}
  
where $S$ is the squeezing, $N^2$ is the stratification near the rim during the advective phase of upwelling and $N^2_0$ is the initial tracer gradient at the same location. Let us consider the deepest isopycnal that upwells onto the shelf $\rho(Z)$ and a density contour above the canyon rim that is mostly unaffected by canyon upwelling, $\rho(h)$ (Fig. \ref{fig:scaling_schematics}). By definition, $\rho(Z)$ is initially at depth $Z+H_h$ and $\rho(h)$ is at depth $h$ with $h<<H_h$ since the effect of the canyon is felt close to the surface (Shallow shelf assumption in AH2010); during the advective phase of upwelling, $\rho(Z)$ rises to approximately depth $H_h$ while $\rho(h)$ stays at $h$. Given these scales, we can approximate $S$ as
\begin{eqnarray}
\label{eq:S}
S& = & \frac{N^2}{N_0^2} \nonumber\\
&\approx& \bigg(\frac{\Delta \rho}{H_h-h}\bigg)\bigg(\frac{\Delta \rho}{H_h+Z-h}\bigg)^{-1} \nonumber\\
& = &1+\frac{Z}{H_h-h} \nonumber\\ 
&\approx & 1+\frac{Z}{H_h}, 
\end{eqnarray}
where $\Delta \rho=\rho(Z)-\rho(h)$. The last step comes from the fact that $h<<H_h$. Additionally, the enhanced, non-uniform stratification will be diffused as a function of time and the local value of $K_v$. For smoother $K_v$ profiles ($\epsilon>5$ m), $K_v$ is larger than $K_{bg}$ above the rim and the effect of diffusion over the enhanced stratification will be larger. We find that the effect of $S$ and the local diffusivity $K_{Z}$ can be expressed as
\begin{equation}
\label{eq:Supw}
S_{upw} = \frac{Z}{H_h}\exp\left(-\frac{K_{Z}t^*}{Z^2}\right),
\end{equation}
where, $K_{Z}$ is the diffusivity $K_v$ evaluated at a distance $Z$ above or below the canyon rim, depending on the region of interest.
 
 If diffusivity within the canyon is high enough that the time scale on which diffusion acts is on the order of the duration of the upwelling event, enhanced $K_{can}$ with respect to the background value $K_{bg}$ will increase the squeezing by further diffusing the density gradient above rim depth and thus decrease it below rim depth (Fig. \ref{fig:scaling_schematics}, lower panel). Consider the case without advection, only diffusion acting on the tracer gradient, and the same linear concentration profile. The top part of the water column, above rim depth, has diffusivity $K_{bg}$ and the bottom part, below rim depth, has diffusivity $K_{can}$, with $K_{bg}<K_{can}$. Right at the rim, the change in concentration is driven by the difference in diffusive fluxes given by $K_{can} \partial \rho / \partial z-K_{bg} \partial \rho / \partial z$. We know that, initially, the density derivatives above and below rim depth are the same ($\partial\rho_i/\partial z$), given the initial conditions we imposed, so the flux from below is larger than the flux from above. The flux mismatch increases the density at the rim. We can estimate the diffusion equation at the rim as
\begin{equation}
\frac{\partial \rho}{\partial t} = \frac{\partial}{\partial z}\bigg[(K_{can}-K_{bg})\frac{\partial \rho_i}{\partial{z}}\bigg].
\label{eq:diffusion}
\end{equation}
Assuming that the changes in density in time are of the same order as the density changes around the rim,  we can approximate (\ref{eq:diffusion}) by evaluating $K_v$ just above and below the rim
\begin{equation}
\frac{\Delta \rho}{\Delta t} \approx \frac{\left[K_v\left(H_r+\frac{dz}{2}\right)-K_v\left(H_r-\frac{dz}{2}\right)\right]}{\Delta z}\frac{\partial \rho_i}{\partial z},
\label{eq:diffusion_approx}
\end{equation}
where $H_r$ is the rim depth and $\Delta z >> dz$. So, after a time $\tau=\Delta t$ and approximating $\partial \rho_i/ \partial z = \Delta \rho/ \Delta z$ a length scale for diffusion $Z_{dif}=\Delta z$ is given by
\begin{equation}
Z_{dif} \approx \left[\left \{ K_v\left(H_r+\frac{dz}{2}\right)-K_v\left(H_r-\frac{dz}{2}\right)\right \} t^*\right]^{1/2}.
\label{eq:Zdif}
\end{equation}
 Physically, $Z_{dif}$ is the initial depth of the isopycnal that reaches rim depth at time $\tau$. Another way to understand $Z_{dif}$ is as the depth that the region with mismatched flux has extended below the rim. In the canyon, advection is the main driver of tracer contour upwelling but, if the difference between $K_{bg}$ and $K_{can}$ is large enough, in just a few days diffusivity can equally contribute to the vertical displacement of isopycnals ($Z_{dif} \approx Z$). 

\begin{figure*}[th!]
\begin{center}
	\noindent\includegraphics[width=0.7\textwidth]{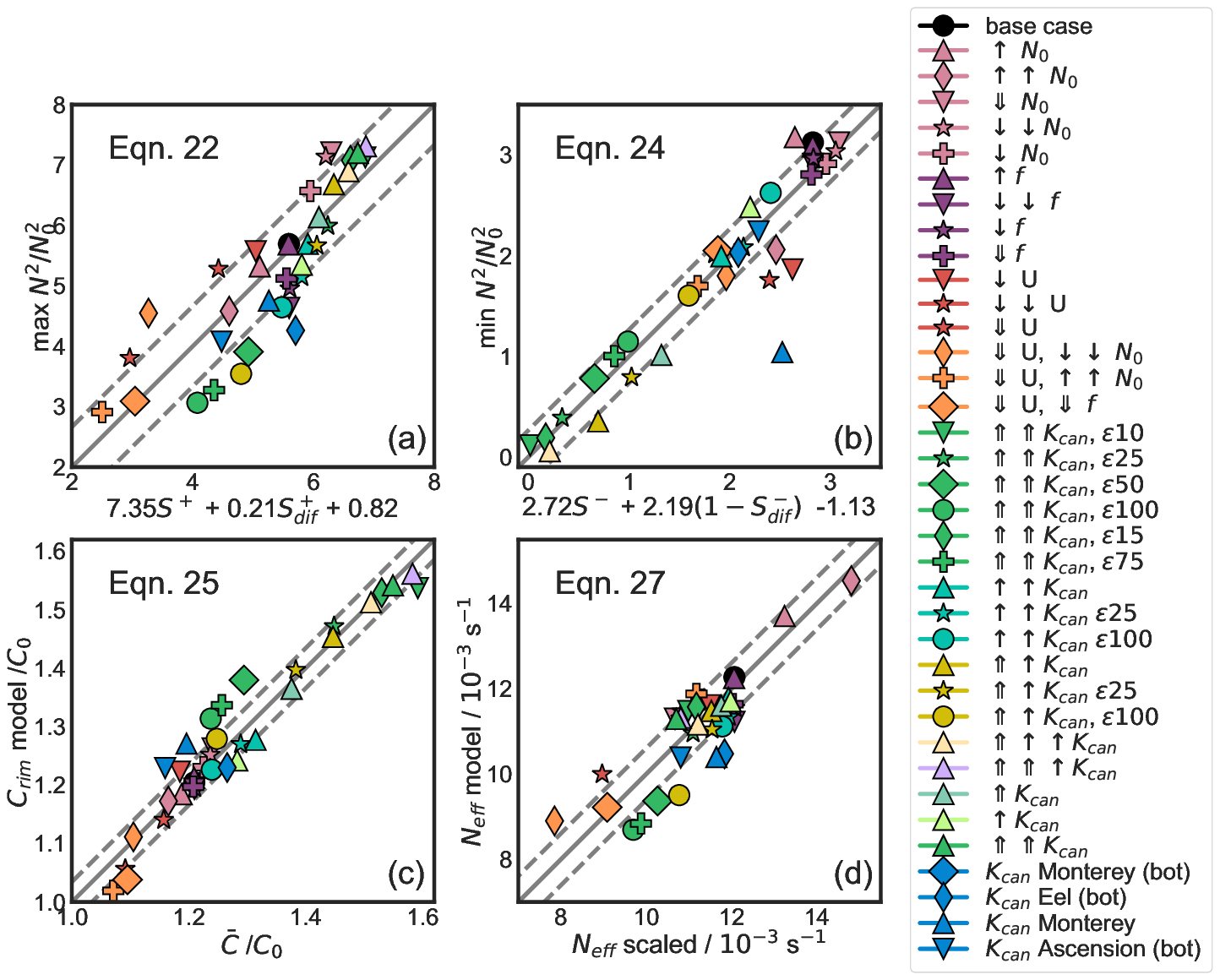}
	\caption{Scaling estimates of maximum stratification $N_{max}$ above the canyon (a), minimum stratification below rim depth $N_{min}$ (b), tracer concentration just above rim depth $H_r$ (c), and effective stratification $N_{eff}=0.75N_{max}+0.25N_{min}$ (d). Dashed lines correspond to $\pm$ one mean squared error.}
	\label{fig:scaling_N_C}
	\end{center}
\end{figure*}

The extra squeezing and stretching effect of enhanced $K_{can}$ is then characterized by the length scale $Z_{dif}$ (Fig. \ref{fig:scaling_schematics}, top panel). Note that if $K_{can}=K_{bg}$ there is no extra diffusion and $Z_{dif}=0$. 
We used a 1D model of diffusion (Appendix B) to find the relationship between $Z_{dif}$ and the stretching of the tracer gradient above the rim, which is the exponential function (Fig. \ref{fig:1Dmodel} g):
\begin{equation}
    S^-_{dif}=\exp\left(\frac{-0.15Z_{dif}}{\Delta z}\right),
 \end{equation}
where $Z_{dif}$ is given by (\ref{eq:Zdif}) with $\Delta z$ the vertical resolution of the 1D model (0.25 m).

The diffusion-driven squeezing below the rim has a similar functional form as the upwelling-driven squeezing (\ref{eq:Supw}), in this case using the diffusion distance $Z_{dif}$, the depth scale $\Delta z$ (5 m for our model) and the length scale $\epsilon$:
\begin{equation}
    S^+_{dif}= \frac{Z_{dif}}{\Delta Z}\exp\left(-\frac{K^+_{Z}t^*}{\epsilon^2}\right),
\end{equation}
where $K^+_{Z}$ is the local diffusivity above the rim, as defined for (\ref{eq:Supw}). Taking into consideration both, the effect of advection (\ref{eq:Supw}) and diffusion the total density squeezing is scaled as
\begin{eqnarray}
\frac{N_{max}^2}{N_0^2}& = & max\left(\frac{N^2}{N^2_0}\right) \nonumber\\
                      &\approx & A_1S_{upw}^++ B_1S^+_{dif} + C_1, 
\label{eq:maxN}
\end{eqnarray}
where $S_{upw}^+$ is a function of $K^+_{Z}$, $A_1$=7.35, $B_1$=0.21 and $C_1$=0.82 are best-fit parameters to a multivariable linear regression. Similarly, the total tracer squeezing is scaled as
\begin{eqnarray}
\frac{\partial_zC_{max}}{\partial_zC_0}& = & max(\frac{\partial_zC}{\partial_zC_0}) \nonumber \\
                      &\approx & A_2S_{upw}^++ B_2S^+_{dif} + C_2, 
\label{eq:maxdTr}
\end{eqnarray}
where $A_2=7.30$, $B_2=0.23$ and $C_2=0.82$. 

The stretching of isopycnals is scaled as
\begin{eqnarray}
\frac{N_{min}^2}{N_0^2}& = & min\left(\frac{N^2}{N^2_0}\right) \nonumber\\
& \approx & A_3S^-_{upw}+ B_3(1-S^-_{dif}) + C_3,
\label{eq:minN}
\end{eqnarray}
 where $A_3=2.72$, $B_3=2.19$ and $C_3=-1.13$. The estimates  compare well with the maximum and minimum stratification (Fig. \ref{fig:scaling_N_C} a and b) and tracer gradient near the canyon head (not shown). 
 
\subsection{Average tracer concentration}
\label{sec:avg_tracer}
The uplift of iso-concentration lines near rim depth provides higher tracer mass on the shelf, especially when $\kappa_{can}$ is enhanced (Fig. \ref{fig:Tr_evolution}g). Thus, we approximate the relative increase in tracer concentration in the vicinity of the rim, just above rim depth, as a function similar to the squeezing of tracer contours: 
 \begin{equation}
 \frac{\bar{C}}{C_0}=A_4S^+_{upw}+B_4S_{dif}^++C_4.
 \label{eq:c_bar}
\end{equation} 
where $C_0$ is the initial concentration at rim depth, and the coefficients $A_4=0.33$ and $B_4=0.06$ and $C_4=1.00$ are proportionality constants. This compares well with the mean concentration near rim depth between days 4 and 9 (Fig. \ref{fig:scaling_N_C}c). For more realistic initial profiles the constant $B_4$ will probably be larger as vertical diffusivity of tracer will have a more prominent role for larger gradients and curvatures in the profiles.

\subsection{Upwelling and tracer fluxes}
\label{sec:upw_flux_scaling}
The scaling estimates by AH2010 state that the dimensionless upwelling flux $\Phi/(W_{m}UD_h)$ is proportional to $\mathcal{F}^{3/2}R_L^{1/2}$, where $D_h=fL/N_0$ is a depth scale. Moreover, HA2013 corrected this estimate to account for the impact of a sloping shelf, since, in a stratified water column, the water upwelled on the continental shelf slope adds pressure that inhibits upwelling, and reduces the upwelling depth and the upwelling flux. Their estimate is
\begin{figure*}[th!]
\begin{center}
	\noindent\includegraphics[width=0.6\textwidth]{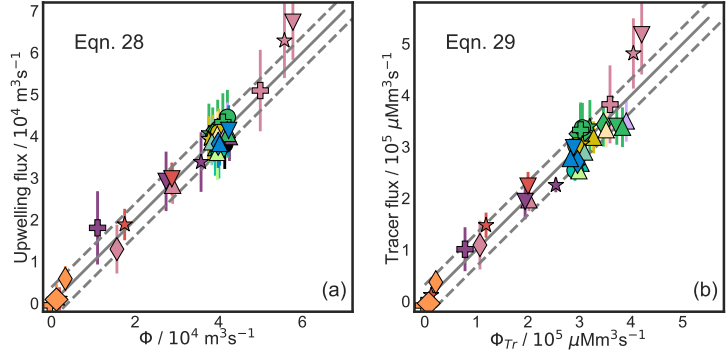}
	\caption{Scaling estimates of upwelling flux of water (a) and upwelling flux of tracer (b) through a submarine canyon. Dashed lines correspond to $\pm$ one mean squared error. Run legend same as for Figure \ref{fig:scaling_N_C}.}
	\label{fig:scaling_PhiTr}
	\end{center}
\end{figure*}

\begin{equation}
\frac{\Phi}{WUD_h} = 0.9\mathcal{F}_w^{3/2}R_L^{1/2}(1-1.21S_E)^3+0.07,
\label{eq:Phi}
\end{equation}
  
where $W$ is the canyon width at mid-canyon length; the function $\mathcal{F}_w=R_W/(0.9+R_W)$ is similar to $\mathcal{F}$ but uses the Rossby number $R_W=U/fW_s$, where $W_s$ is the width at mid-length measured at shelf-break depth. The slope effect is encapsulated in the function $S_E=sN_0/f(\mathcal{F}_w/R_L)^{1/2}$, where $s$ is the shelf slope ($s$=$1\times10^{-2}$ for all runs here). 

We found that locally-enhanced diffusivity has an effect on the upwelling flux: lower stratification in the canyon allows more water to upwell while high stratification above rim depth acts like a `lid' to suppress the upwelling. We propose to use an effective stratification $N_{eff}$ as the scale for stratification in (\ref{eq:Phi}) to account for the effect of enhanced $K_{can}$ and $\epsilon$ where $N_{eff}$ is defined as 
\begin{equation}
N_{eff}=(0.75N_{max}+0.25N_{min}), 
\label{eq:Neff}
\end{equation}
where $N_{max}$ and $N_{min}$ are the maximum (\ref{eq:maxN}) and minimum (\ref{eq:minN}) stratification above and below rim depth, respectively (Fig.\ref{fig:scaling_N_C}d). This gives
 
\begin{equation}
\frac{\Phi}{WUD_{eff}} = 4.98\mathcal{F}_w^{3/2}R_L^{1/2}(1-0.52S_E)^3-0.01,
\label{eq:Phi_RA}
\end{equation}
  
 where $D_{eff}=fLN_{eff}^{-1}$ and the coefficients were re-fitted to satisfy the equation. Note that $N_{eff}$ is only used to calculate the depth scale $D_{eff}$.  This estimate compares well with the mean upwelling flux calculated from days 4 to 9 (Fig. \ref{fig:scaling_PhiTr}a and Tables S2). Upwelling flux increases by approximately 19\% with respect to the base case for the largest $K_{can}$ case (Table \ref{tab:HCW_TrMass}, column 2).

In section 3\ref{sec:HCW} we found that the tracer flux upwelled onto the shelf by the canyon is directly proportional the upwelled water flux. Consequently, we approximate the total upwelled tracer flux as the product of the upwelling flux (\ref{eq:Phi_RA}) and the average tracer concentration near rim depth within the canyon (\ref{eq:c_bar}):
 
\begin{equation}
\Phi_{Tr}=A_5\bar{C} \Phi+B_5.
\end{equation}
  
 where $A_5 = 1.00$ and $B_5=-718.86$ $\mu$Mm$^3$s$^{-1}$ are best-fit, least-square parameters. This estimate compares well with the mean upwelled tracer flux calculated from days 4 to 9 shown in column 3 of Table S2 (Fig. \ref{fig:scaling_PhiTr}b). The relatively larger concentration near the canyon rim characterizes the increased tracer mass flux when vertical diffusivity is enhanced locally (27\% for the largest $K_{can}$ case) while the lower $N_{min}$ enhances the upwelling flux of water. Our scaling estimate successfully quantifies these effects.
 
We included five runs with $K_{v}$ profiles inspired in observations to provide context to our scaling (Fig. \ref{fig:scaling_PhiTr} a and b, blue markers). Our scaling works well when using less idealized profiles, but it cannot be applied to profiles with $K_{can}<K_{bg}$ because the scale $Z_{dif}$ is not defined. Nonetheless, these cases prove that our scaling is robust enough to work with non-smooth profiles as the ones that could be measured in a canyon. Additional smoothing of measured profiles could be done to apply our scaling. See SI, Figure S2 for the methodology followed to develop these runs.

\section{Discussion and conclusions}
\label{sec:Discussion}
Advection-induced upwelling of water through a canyon is the dominant driver of on-shelf transport of tracer mass from the open ocean, however, the tracer concentration profile and enhanced vertical diffusivity within the canyon contribute considerably to the amount and spatial distribution of the tracer on shelf. The main characteristics of canyon-induced tracer upwelling are the following (Fig. \ref{fig:Adv_phase}):
\begin{figure*}[th!]
\begin{center}
   		\noindent \includegraphics[width=0.8\textwidth]{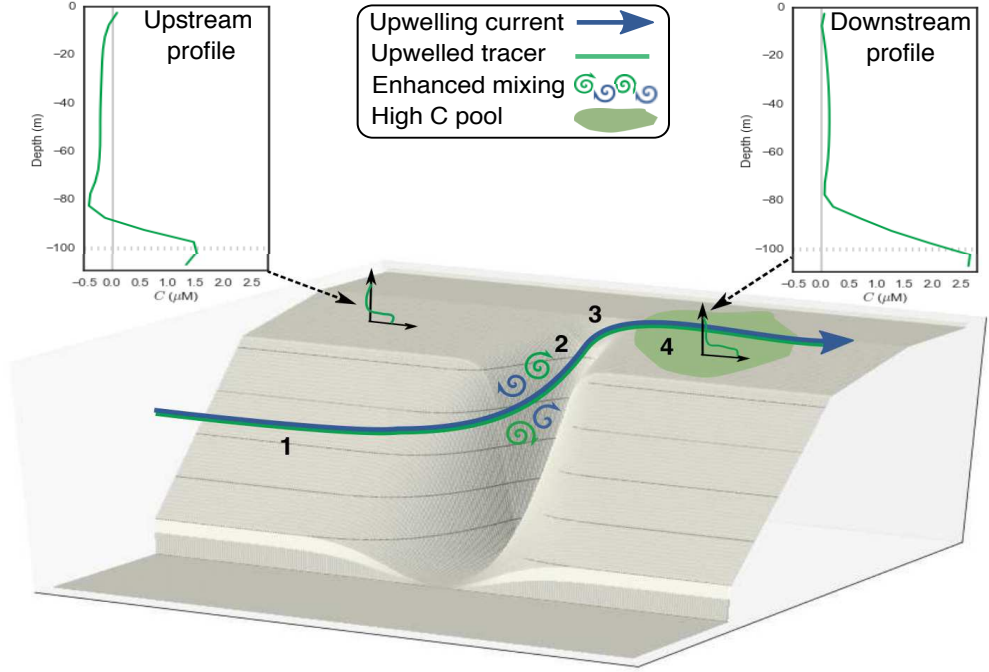} 
		\caption{Schematics of tracer transport through a submarine canyon: 1) The upwelling current (blue arrow) brings tracer-rich water onto the shelf, generating an area of relatively higher tracer concentration than the upstream shelf (4). Enhanced vertical diffusivity within the canyon (2 and 3) increases the tracer concentration near rim depth and weakens the stratification. These two effects enhance canyon-induced tracer flux onto the shelf.}
		\label{fig:Adv_phase}
		\end{center}
\end{figure*}
\begin{enumerate}
\item{The upwelling flux carries tracer onto the shelf near the head and the downstream side of the canyon rim, to be further spread on the shelf; with decreasing $B_u$ and increasing $R_W$, the amount transported is larger. Also, for a tracer profile that increases with depth, a larger upwelling depth will bring water with higher concentration onto the shelf; with decreasing $B_u$ and increasing $R_W$, the depth of upwelling is larger.}
\item{Locally-enhanced mixing weakens the stratification below rim depth. A smaller stratification increases the vertical advective transport of water and thus, of tracers. The mechanism is that isopycnals close to the head are squeezed due to upwelling, which generates a local increase in stratification proportional to the isopycnal tilting generated by upwelling. However, enhanced diffusivity acts against temperature and salinity gradients, thus reducing this density gradient and locally reducing stratification below the rim. The combined effect of lower $N$ and higher diffusivity below the rim via a smoother $K_v$ profile (larger $\epsilon$) can increase the water flux by up to 26\% for values chosen in this study.}
\item{Enhanced mixing within the canyon increases the tracer concentration near rim depth. Just above rim depth, where the value of $K_{can}$ changes, the tracer gradient increases. This means that concentration isolines are elevated higher compared to the situation with uniform diffusivity and in turn, isolines of higher concentrations will be reaching rim depth. This water with higher tracer concentration will upwell. Together, this mechanism and 2 above increase the tracer flux onto the shelf. For instance, taken together both contributions can increase tracer upwelling flux by 27\% when $K_{can}$ is locally enhanced by three orders of magnitude.}
\item{The upwelled water spreads out on the shelf, downstream of the rim and generates a region of relatively larger tracer concentration near the bottom.}
\end{enumerate}

For comparison,  \cite{Messie2009} estimated that the wind-driven nitrate supply for the Northern Washington Shelf is 6.4~mmol~s$^{-1}$m$^{-1}$. This corresponds to 153~mol~s$^{-1}$ across a shelf section of length $W_m=24$~km, the width of the canyon, while the tracer upwelled through the canyon (VTT) for the base case is 160~mol s$^{-1}$. Considering that the `nitrate' concentration of the upwelled water in our model is about 4 times smaller than it would be in a coastal environment like the West Coast of Vancouver Island, then the canyon supplies 4 times more nitrate than wind-driven upwelling. For a typical $K_{can}$ profile, enhanced diffusivity increases the transport by 25\%, thus increasing the transport by an amount similar to wind-driven upwelling.


%
\subsection{Implications on Internal Waves}
Enhanced, upwelling-induced stratification near rim depth observed in our numerical results can potentially alter the propagating characteristics of internal waves in the canyon by two mechanisms. First, canyons are known to focus internal waves towards the canyon floor. Their wedge-shaped topography is supercritical to the most energetic type of internal waves found on the nearby-shelf \citep{Gordon1976}. Enhanced stratification near rim depth, close to the canyon head (upper canyon), can increase the criticality,  $\alpha$, of the upper canyon walls given that it is dependent on the buoyancy frequency $N$:
\begin{equation}
\alpha = \frac{S_{topo}}{S_{wave}}=\frac{\partial H/ \partial x}{[(\omega^2-f^2)/(N^2-\omega^2)]^{1/2}},
\end{equation}
where $S_{topo}$ is the topographic slope, $S_{wave}$ is the wave characteristic slope, $x$ is the cross-slope direction, $H$ is the total water depth, $\omega$ is the wave frequency, $f$ the Coriolis parameter and $N$ the buoyancy frequency. A slope is supercritical when $\alpha > 1$ and will reflect the wave towards deeper water which can mean towards the canyon floor if the incident wave is perpendicular to the canyon walls, and down-canyon if the incident wave is perpendicular to the canyon axis. 

The second mechanism is the transition between a partly standing wave during pre-upwelling conditions to propagating during upwelling conditions. This effect has been observed \citep{Zhao2012} and modelled \citep{Hall2014} for the $M_2$, mode 1 internal tide in Monterey Canyon. During pre-upwelling conditions, the pycnocline was located below rim depth, which increased the supercritical reflections (down-canyon) of the up-canyon propagating internal tide. During upwelling conditions, the pycnocline rose above rim depth, decreasing the stratification and with it, the supercriticality of the canyon walls. This decreased stratification decreased the reflection of the up-canyon propagating tide. The comparatively large reflection during pre-upwelling conditions allowed for a horizontally, partly-standing wave set up, while upwelling conditions caused a progressive up-canyon wave to dominate.

In our model, maximum stratification within the canyon and near the rim is a consequence of shelf-break and canyon-induced upwelling where isopycnals tilt towards the canyon head, squeezing closer together around rim depth, not too far above the canyon walls. This enhanced stratification could push the reflecting characteristics of the canyon walls or bottom towards the supercritical regime as results from \cite{Zhao2012} and \cite{Hall2014} suggest. Moreover, our results show that having elevated diffusivity within the canyon will erode the increased, canyon-induced stratification below rim depth and enhance it above rim depth. If we assume that the stratification that matters for criticality occurs around rim depth, then the competition between squeezing and stratification erosion will determine the change in criticality. Close above the rim we see stratification ($N^2$) increasing up to 5 times due to canyon-induced upwelling and up to 7.5 times when $K_v$ is locally-enhanced, which could translate in a change in $\alpha$ from 0.4 to 0.8 - 1.0 (for $5N_0^2$ and $7.5N_0^2$, respectively) alongshelf, and from 1.4 to 3.1-3.9 near the canyon head along the axis. Below rim depth, enhanced diffusivity can erode the isopycnal squeezing to be 0.3$N_0^2$, decreasing the maximum value of $\alpha$ alongshelf from supercritical to subcritical (1.4 to 0.8).

Upwelling in short canyons is stronger on the downstream half of the canyon and thus, the eroding effect of enhanced diffusivity over increased stratification will also be stronger there due to the large upwelling-generated gradients. So, the change in criticality will be impacted by this asymmetry too. A larger shift towards supercriticality is to be expected on the downstream side of the canyon, close to the head and strongly modulated by the difference in diffusivity below and above rim depth. This shift will also influence the location of internal wave breaking and, as a consequence, where vertical diffusivity is enhanced.

\subsection{Extension to other canyons}
The diffusivity-driven weakening of vertical gradients is a function of time. There is a natural time scale in which diffusivity acts on vertical gradients given a characteristic length scale, for example, the upwelling depth. The larger the diffusivity the smaller the time scale given the same length scale. We find that diffusivities of around O($10^{-3}$ m$^2$s$^{-1}$) or above are sufficiently high to noticeably weaken stratification and tracer gradient in the first 4 days. This means that when the flow enters the advective phase, the effects of high $K_{can}$ are already noticeable. Enhanced diffusivity continues to act on the gradients during the advective phase but the effect weakens as it is proportional to the gradient itself. In canyons such as Monterey, where diffusivities are on the order of $10^{-2}$ m$^2$s$^{-1}$, the weakened gradients would be considerable after only 11 hours, assuming a depth of upwelling of about 20 m.  

Our results and overall scaling scheme are valid only for short canyons, which are canyons for which the canyon head occurs well before the coast \citep{Allen2000}. This criterion removes some of the most iconic canyons, like Monterey and Nazar\'e Canyons. For canyons not in the \cite{Allen2010} scaling, we expect that, provided there is squeezing of isopycnals and a difference in diffusivity above and below the rim, the same effect of non-uniform diffusivity would occur: the differentiated diffusivity will act to further enhance the stratification above the rim and further decrease it below the rim. The tracer part of the scaling would be similar but an appropriate depth of upwelling, $Z$, and fitting parameters would need to be found. For less idealized bathymetries the overall upwelling pattern is expected to be very similar, provided that the incoming flow is along the shelf, perpendicular to the canyon axis, and relatively uniform along the length of the canyon. Scaling of the upwelling flux and depth of upwelling is robust enough that it has been successfully applied to real, short canyons like Astoria, Barkley and Quinault Canyons (AH2010) and in one of the limbs of Whittard Canyon (depth of upwelling in \cite{Porter2016}).

Runs with longer canyons (2 times and 1.5 times longer than our original canyon) show that the general circulation pattern and evolution of the upwelling event is similar, as seen in \cite{Howatt2013}. Isopycnals and iso-concentration lines tilt towards the canyon head similarly for both canyons, so that squeezing of isopycnals happens close to the head in both cases. The stratification evolution near canyon head, on the downstream side of the canyon is also similar for longer canyons. Moreover, having locally-enhanced diffusivity within the canyons has the same effect on isopycnal squeezing near the canyon head. Locally-enhanced diffusivity increases the near-rim depth concentration in all three cases compared to the case with uniform diffusivity and the concentration is well predicted by (25) with root mean square error 0.04 compared to 0.03 for the single canyon. These runs show that the effect of diffusivity can be applied to other canyons, whenever there is isopycnal squeezing and different diffusivities above and below the rim.

The tracer mass flux scaling estimated in this work is restricted to flows that follow the same conditions as AH2010 and HA2013 because it depends on their upwelling flux estimation and as such, it can only perform as good as their estimate. The main contribution of our scaling scheme is the estimation of tracer concentration and stratification within the canyon. Our scaling preformed reasonably well when we used it on runs with $K_v$ profiles inspired by observations.  

\subsection{Significance to upwelling nutrients}
\cite{Connolly2014} identified a similar feature to the pool. They estimated that canyon-exported nitrate onto the shelf after two months during an upwelling season can be about 1-2$\times10^{7}$ kg NO$^-_3$. We found that after a single upwelling event (9 days) the canyon can increase the total inventory of tracer mass on the shelf by 0.3-4.7$\times10^6$ kg NO$^-_3$ compared to a straight shelf case. If we consider a 60 day upwelling period, then the canyon contribution to the tracer inventory could be up to 3$\times10^7$ kg NO$^-_3$. Additionally, after a canyon upwelling event, between 24 to 89\% of the upwelled tracer mass on the shelf can be canyon upwelled, given a canyon with a width that represents about 5\% of the shelf length and depending on the dynamical characteristics of the flow. 

Future work will consider scaling for realistic profiles of nutrients and oxygen as well as characterizing the pool of upwelled water and tracers that forms on the downstream shelf. Some of the key features to consider are the slope and curvature of the profile as suggested by the scaling of the advection-diffusion equation, and the location of the nutricline and oxygen-minimum zone.

%
\acknowledgments
The authors would like to thank J. Klymak and S. Waterman for sharing insightful comments about the project, A. Waterhouse and G. Carter for providing measurements of diffusivity profiles, and D. Sheinbaum for fruitful discussions. Computing power was provided by WestGrid and Compute Canada. This work was funded by NSERC Discovery Grant RGPIN-2016-03865 to SEA and UBC through a Four Year Fellowship to KRM. The model configuration and post processing scripts can be consulted from our repository at https://bitbucket.org/canyonsubc/tracer\_upwelling\_paper.

%






%
%
%
\newpage
\appendix[A]
\appendixtitle{Advection-diffusion equation in natural coordinates}
We need to express the rhs of (\ref{eq:adv-diff_tau_eta_b}) in terms of ($s$,$z$,$n$) to compare the relative size of each term. To do that we calculate the first and second spatial derivatives of the concentration:
 
 \begin{equation}
 \frac{\partial{C}}{\partial{\eta}}=\frac{\partial{C}}{\partial{s}}\frac{\partial{s}}{\partial{\eta}}+\frac{\partial{C}}{\partial{n}}\frac{\partial{n}}{\partial{\eta}}+\frac{\partial{C}}{\partial{z}}\frac{\partial{z}}{\partial{\eta}}.
 \end{equation}
   
Note that,
 
\begin{align}
 \frac{\partial{n}}{\partial{\eta}}&=\frac{\partial{n}}{\partial{\tau}}=0, \label{eq:approxdndtau}\\
 \frac{\partial{z}}{\partial{s}}&=\frac{\partial{n}}{\partial{s}}=0,
 \label{eq:approxdnds}
\end{align}
  
 and
 
  \begin{equation}
 \frac{\partial{n}}{\partial{b}}=1,
 \label{eq:dndeta_is_zero}
 \end{equation}
   
since $n=b$. Further, 
   
  \begin{align}
  \frac{\partial{z}}{\partial{\eta}}\approx\frac{\partial{s}}{\partial{\tau}}\approx\cos{\alpha},\\
 \frac{\partial{s}}{\partial{\eta}}\approx\frac{\partial{z}}{\partial{\tau}}\approx\sin{\alpha},
\end{align}
  
so that for small angles,
 
 \begin{align}
 \cos{\alpha}\approx1,\\
 \sin{\alpha}\ll1.
 \end{align}
   
The second derivative with respect to $\eta$, after eliminating terms and approximating the trigonometric functions of small angles is
  
 \begin{equation}
 \frac{\partial}{\partial{\eta}}\left(\frac{\partial{C}}{\partial{\eta}}\right)=\frac{\partial}{\partial{\eta}} \left(\frac{\partial{C}}{\partial{s}}\frac{\partial{s}}{\partial{\eta}}\right)+\frac{\partial}{\partial{\eta}}\left(\frac{\partial{C}}{\partial{z}}\frac{\partial{z}}{\partial{\eta}}\right).
\end{equation}
  
Expanding and eliminating terms according to (\ref{eq:approxdnds}) and (\ref{eq:dndeta_is_zero}) gives
 
\begin{equation}
 \frac{\partial^2C}{\partial{\eta^2}} \approx 2\frac{\partial^2C}{\partial{z}\partial{s}}\frac{\partial{s}}{\partial{\eta}}\frac{\partial{z}}{\partial{\eta}}+ \frac{\partial^2C}{\partial{z^2}}\left(\frac{\partial{z}}{\partial{\eta}}\right)^2.
 \end{equation}
   
The second derivative with respect to $\tau$ is approximated as
 
 \begin{equation}
 \frac{\partial^2C}{\partial{\tau^2}} \approx 2\frac{\partial^2C}{\partial{z}\partial{s}}\frac{\partial{s}}{\partial{\tau}}\frac{\partial{z}}{\partial{\tau}}+ \frac{\partial^2C}{\partial{s^2}}\left(\frac{\partial{s}}{\partial{\tau}}\right)^2.
 \end{equation}
   
 Finally, the second derivative with respect to $b$ is
 
\begin{equation}
 \frac{\partial^2C}{\partial{b^2}}=\frac{\partial^2{C}}{\partial{n^2}}.
 \end{equation}
   
The final approximation of  (\ref{eq:adv-diff_snz}) is
  
  \begin{align}
  \frac{\partial{C}}{\partial{t}}+u\frac{\partial{C}}{\partial{s}}+w\frac{\partial{C}}{\partial{z}}&\approx K_{I}\left(\frac{\partial^2C}{\partial{s^2}}+\frac{\partial^2C}{\partial{n^2}}+2\frac{\partial^2C}{\partial{z}\partial{s}}\frac{\partial{z}}{\partial{\tau}}\right) \nonumber\\
  &+ K_{D}\left(\frac{\partial^2C}{\partial{z^2}}+2\frac{\partial^2C}{\partial{z}\partial{s}}\frac{\partial{s}}{\partial{\eta}}\right). 
  \label{eq:adv-diff_natural_app}
  \end{align}
  
\appendix[B]
\appendixtitle{1D model of diffusion}
\section{Appendix B}
\begin{figure*}[th!]
\begin{center}
   		\noindent \includegraphics[width=0.8\textwidth]{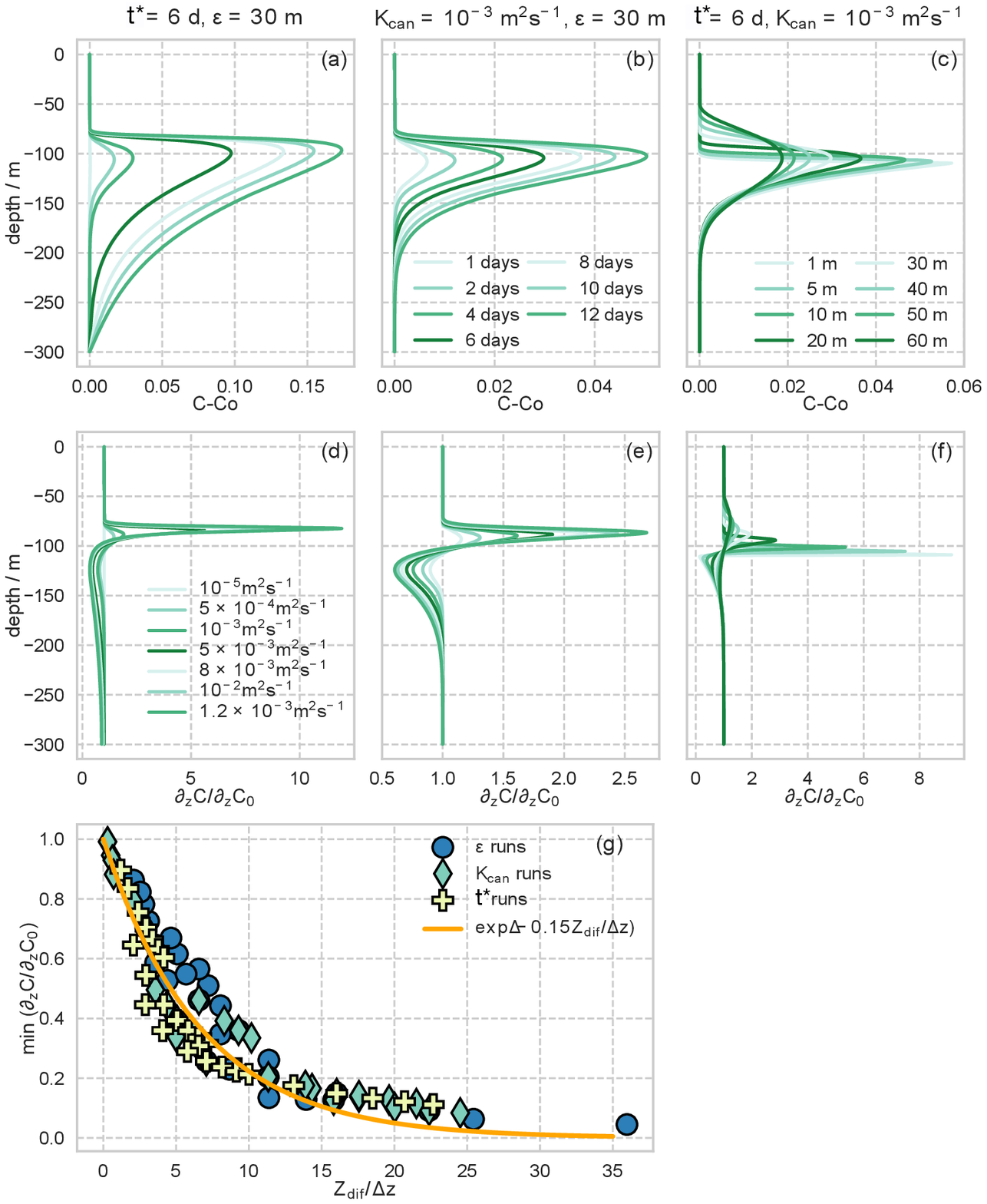} 
		\appendcaption{B1}{(a-c) Tracer concentration difference from the initial profile for runs from the 1D diffusion model varying (a) $K_{can}$, (b) $\tau$ and (c) $\epsilon$. (d-f) Corresponding tracer profile gradients. (g) Minimum tracer gradient for 1D model runs covering the parameter space $\epsilon=1$ to 50 m ($\epsilon$ runs), $K_{can}=10^{-5}$ to $1.2\times10^{-2}$ ($K_{can}$ runs) and $\tau=1$ to 12 days ($\tau$ runs). The orange line corresponds to the fitted decreasing exponential function relating the stretching and $Z_{dif}/\Delta z$.}
		\label{fig:1Dmodel}
		\end{center}
\end{figure*}

We use a 1D model of diffusion through two layers of water with different diffusivities to illustrate the effect of a sharp diffusivity profile and progressively smoother versions of that step described by the smooth Heaviside function (\ref{eq:heaviside}). Increasing $\epsilon$ increases the depth where the concentration is changing due to a mismatch in the flux (Fig. \ref{fig:1Dmodel}, panels a-c), and at the interface ($z=H_r$) we see a smaller increase in concentration relative to the step profile. 


\bibliographystyle{ametsoc2014}
\bibliography{tracer_transport_canyon}

\end{document}